\documentclass[twocolumn]{aa}
\usepackage{amsmath}
\usepackage{mathrsfs}
\usepackage{graphicx}
\usepackage{epsfig}
\usepackage{pifont}
\usepackage{color}
\usepackage{txfonts}
\usepackage{multirow}
\definecolor{turq}{cmyk}{1,0,0,0.5}

\DeclareMathAlphabet{\mathpzc}{OT1}{pzc}{m}{it}

\begin{document}

\title{Coronal heating by the partial relaxation of twisted loops}

\author{M. R. Bareford\inst{1}\and A. W. Hood\inst{1}\and P. K. Browning\inst{2}}
\offprints{M. R. Bareford} \institute{School of Mathematics and Statistics, University of St. Andrews, North Haugh, St. Andrews, Fife KY16 9SS, UK\\ \email{michaelb@mcs.st-and.ac.uk} \and Jodrell Bank Centre for Astrophysics,
Alan Turing Building, School of Physics and Astronomy, University of Manchester, Oxford Road, Manchester M13 9PL, UK}
\date{Received 04/10/2012; accepted 01/11/2012}

\abstract
% context heading (optional)
{Relaxation theory offers a straightforward method for estimating the energy that is released when continual convective driving causes a magnetic field to become unstable. Thus, an upper limit to the heating caused by ensembles of nanoflaring coronal loops can be calculated and checked against the level of heating required to maintain observed coronal temperatures ($T\,{\gtrsim}\,10^6\,\mathrm{K}$).}
% aims heading (mandatory)
{We present new results obtained from nonlinear magnetohydrodynamic (MHD) simulations of idealised coronal loops. All of the initial loop configurations discussed are known to be linearly kink unstable. The purpose of this work is to determine whether or not the simulation results agree with Taylor relaxation, which will require a modified version of relaxation theory applicable to unbounded field configurations. In addition, we show for two cases how the relaxation process unfolds.}
% methods heading (mandatory)
{A three-dimensional (3D) MHD Lagrangian-remap code is used to simulate the evolution of a line-tied cylindrical coronal loop model. This model comprises three concentric layers surrounded by a potential envelope; hence, being twisted locally, each loop configuration is distinguished by a piecewise-constant current profile, featuring three parameters. Initially, all configurations carry zero-net-current fields and are in ideally unstable equilibrium. The simulation results are compared with the predictions of helicity-conserving relaxation theory.}
% results heading (mandatory)
{For all simulations, the change in helicity is no more than 2\% of the initial value; also, the numerical helicities match the analytically-determined values. Magnetic energy dissipation predominantly occurs via shock heating associated with magnetic reconnection in distributed current sheets. The energy release and final field profiles produced by the numerical simulations are in agreement with the predictions given by a new model of \textit{partial} relaxation theory: the relaxed field is close to a linear force free state; however, the extent of the relaxation region is limited, while the loop undergoes some radial expansion.} 
% conclusions heading (optional)
{The results presented here support the use of partial relaxation theory, specifically, when calculating the heating-event distributions produced by ensembles of kink-unstable loops. The energy release increases with relaxation radius; but, once the loop has expanded by more than $50\%$, further expansion yields little more energy. We conclude that the relaxation methodology may be used for coronal heating studies.}

{}

\titlerunning{Coronal heating by partial relaxation}
\maketitle

\section{Introduction}
\label{sec_Introduction}
The energy required to heat the solar corona is thought to originate from the magnetic fields that permeate the Sun's atmosphere. The geometry of these fields is revealed by coronal loops, where the emitting plasma is constrained to follow the magnetic field: the plasma beta is extremely low ($\beta\,{\approx}\,0.01$). Coronal loops are closed structures that emerge from the photosphere at one location and re-enter the solar surface at another. The convective motions at these photospheric boundaries (or footpoints) are thought to reconfigure coronal fields and thereby cause free magnetic energy to accumulate, making the fields more susceptible to instability. Essentially, the kinetic energy of the convection zone is transported, via a Poynting flux, through the photosphere and stored within the coronal loop. In the case of a single loop, currents are created by those convective motions that cause the footpoints to be twisted. The approximate time scale for the twisting motions is long compared to the Alfv\'en time, and so the coronal loop (not necessarily illuminated) transitions through a series of force-free equilibria, which can be expressed as $\nabla\times\vec{B}\,{=}\,\alpha(\vec{r})\vec{B}$; where \textit{$\vec{r}$} is a position vector, and $\alpha\,{=}\,(\mu_{0}\vec{j}\cdot\vec{B})$/$|\vec{B}|^{2}$ is related to the parallel electric current density (i.e., $\alpha$ is a measure of the twist).

Energy release might be triggered when a loop-like magnetic field, driven by continual convective motions, reaches the threshold for kink instability (Hood 1992; Browning \& Van der Linden 2003, Haynes \& Arber 2007; Srivastava et al. 2010). Coronal magnetic fields cover many thousands of kilometers ($\mathpzc{L}\,{\approx}\,50\,\mathrm{Mm}$) and exist within a highly conductive environment: therefore, in the absence of instability, magnetic fields diffuse slowly ($t_{\mathrm{d}}\,{\approx}\,10^3\,\mathrm{yr}$). Nevertheless, through an ideal kink instability, a coronal loop may be deformed such that magnetic flux surfaces are brought together. As the deformation continues, areas of high current are produced, allowing magnetic reconnection to take place. Hence, energy is released from the magnetic field during the nonlinear phase of an ideal kink instability, as shown by many three-dimensional (3D) magnetohydrodynamic (MHD) models (Baty \& Heyvaerts 1996; Velli et al. 1997; Arber et al. 1999; Baty 2000; Browning et al. 2008; Hood et al. 2009). 

The coronal heating problem is most pronounced within active regions, where the heating requirement is approximately $10^{7}\,\mathrm{erg}\,\mathrm{cm}^{-2}\,\mathrm{s}^{-1}$, significantly higher than that necessary for the quiet Sun (Withbroe \& Noyes 1977; Aschwanden \& Acton 2001). These regions are often observed as dense thickets of transient coronal loops, the composition of which changes over the course of hours to days. Sudden reconfigurations are coincident with large-scale solar flares (${\sim}\,10^{34}\,\mathrm{erg}$). The signal strength of such phenomena means these events are more readily observed --- detailed investigations have revealed strong evidence for magnetic reconnection (Fletcher 2009; Qiu 2009). Miniscule (nano)flares, far weaker than ones commonly observed, could maintain coronal temperatures if these less dramatic events occur with sufficient frequency (Hudson 1991). This type of flaring might be the sort initiated by the kink instability; however, the amount of energy released (i.e., the difference between the energy at instability onset and that of the relaxed field) depends on how much the magnetic field has altered before it relaxes. Bareford et al. (2010, 2011) identified the threshold for linear kink instability with respect to an idealised coronal loop model (both with and without net current). This work determined the subset of field configurations accessible via convective driving that are linearly kink unstable. One approach to calculating the energy release, is to represent each of these configurations within a nonlinear MHD code, allow the instability to take place and follow the reconnection dynamics until a relaxed state emerges. However, it is simply not feasible to run such a computationally-intensive process for more than a few examples. Hence, Bareford et al. (2010, 2011), used relaxation theory to identify the relaxed states for all of the threshold (i.e., marginally unstable) configurations.

An unstable field obeys relaxation theory if it relaxes towards the \textit{lowest} energy state that conserves total magnetic axial flux and \textit{global} magnetic helicity (Taylor 1974, 1986). This minimum energy state is a linear force-free field; i.e., $\alpha(r)$ is invariant and $\nabla\times\vec{B}\,{=}\,\alpha\vec{B}$. The helicity ($K$) indicates how intertwined a magnetic field is with itself (Berger 1999). Coronal loops have a non-zero normal flux at the footpoints; hence, the gauge-invariance of relative helicity (Berger \& Field 1984; Finn \& Antonsen 1985) makes it the more useful property:
\begin{eqnarray}
  \label{relative_helicity}
  K & = & \int_V (\vec{A}+\vec{A'})\cdot (\vec{B}-\vec{B\,'})\,\,dV,
\end{eqnarray}
where $\vec{A}$ is the magnetic potential, $\vec{B}'$ is the potential field with the same boundary conditions and $\vec{A}'$ is the corresponding vector potential. Helicity-conserving relaxation has been seen in many laboratory experiments (Heidbrink \& Dang 2000; Taylor 1986). It must be noted that helicity is subject to global resistive diffusion. Nevertheless, if localised dissipation occurs on small spatial scales (i.e., across shock fronts or within thin current sheets), the reduction in helicity will be negligible compared to the decrease in magnetic energy (Browning 1988, Browning et al. 2008). The original intention of relaxation theory was to explain laboratory plasma phenomena; but latterly, it has been frequently applied to the solar corona (Heyvaerts \& Priest 1984; Browning et al. 1986; Vekstein et al. 1993; Zhang \& Low 2003; Priest et al. 2005).

The relative ease with which relaxation theory can be applied meant that Bareford et al. (2010, 2011) were able to generate heating-event distributions from ensembles of idealised coronal loops, representing, albeit crudely, the population of coronal loops that exist within an active region. Each energy release is determined by where a loop crosses the instability threshold; this location is the outcome of a defined stochastic process. The distributions lead to an estimate for the heating rate that is just sufficient for coronal heating. However, the assumptions of this work regarding instability and relaxation theory have yet to be tested by a nonlinear 3D MHD code. The purpose of this paper is to elucidate further (Browning et al. 2008; Hood et al. 2009) the relaxation process and to understand how it can be applied to coronal loops that lack a conducting wall. We simulate a set of zero-net-current coronal loops that sample the linear instability threshold calculated by Bareford et al. (2011). This is a larger set than the one investigated by Hood et al. (2009), and futhermore, the loops represented here feature a current-neutralisation layer that maintains zero net current even if the currents inside the loop are predominantly single signed. (Loops that carry zero net current are preferred since the convective motions that twist the loop and thereby create azimuthal field are spatially localised; the field outside the loop is unaffected by motions within the loop cross-section and therefore remains purely axial.)

A long-standing problem has been how to apply relaxation theory in astrophysical contexts without the presence of conducting walls: simplistically, the relaxation should extend out to infinity and lead always to potential fields. Browning (1988) and Dixon et al. (1989) showed that relaxation theory could apply to volumes with free boundaries, but did not give a prediction for the spatial extent of the relaxed state. We find that a modification to Taylor relaxation theory is required before it can be used to estimate the energy released by a kink-unstable loop. In contrast to previous work, we calculate the helicity directly at specific times during each simulation (Browning et al. 2008 integrated the time differential of the helicity in order to show the change in $\delta K$); thus, we are able to verify the extent of helicity conservation. We also examine the performance of the MHD code with regard to energy conservation. Any numerical dissipation will have implications for the modelling of plasma processes associated with heating, such as radiation. However, the plasma-$\beta$ is sufficiently low that any artificial resistivity should not influence the energy released by an unstable magnetic field, nor should it affect the evolution of a relaxing field.

The paper is structured in the following manner. Section \ref{sec_NumericalCode} describes the numerical code used for the nonlinear MHD simulations, along with the loop model and the equations used to calculate the magnetic field. The corresponding instability threshold for the linear kink mode is introduced, as are the positions of the simulated loop configurations. Section \ref{sec_NumericalResults} presents the results: specifically, how the different forms of energy vary over the course of the simulation, when the loop goes unstable, and how these results are affected by changes in the code parameters, such as spatial resolution and background resistivity. Following, the evolution of the loop is presented as regards magnetic field and current magnitude. Section \ref{sec_PartialRelaxationModel} discusses how well the results fit a modified Taylor relaxation theory. Finally, in the last section, the results are summarised and our conclusions are given.

\section{Numerical code}
\label{sec_NumericalCode}
The nonlinear simulations are conducted using a 3D MHD Lagrangian Remap Cartesian code, called LARE3D (Arber et al. 2001). It is written in Fortran 90 and uses the Message Passing Interface (MPI) to achieve parallelisation. The Lagrangian step uses a second-order accurate predictor-corrector step that also incorporates artificial viscosity, ensuring shocks are captured accurately. Van Leer (1997) gradient limiters are used at the remap step in order to preserve monotonicity. The divergence-free condition ($\nabla\cdot\vec{B}\,{=}\,0$) is maintained to machine precision by Evans \& Hawley (1988) constrained transport.

LARE3D solves the resistive MHD equations given by
\begin{eqnarray}
  \label{eqn_lare_mhd_mass}
  \frac{\partial\rho}{\partial t} & = & -\nabla\cdot(\,\rho\vec{v}\,)\,,\\
  \nonumber \\ 
  \label{eqn_lare_mhd_force}
  \frac{\partial}{\partial t}\big(\,\rho\vec{v}\,\big) & = & -\nabla\cdot(\,\rho\vec{v}\vec{v}\,)\,\,+\,\,\frac{1}{\mu_0}\Big(\,\nabla\times\vec{B}\,\Big)\times\vec{B}\,\,-\,\,\nabla{P}\,\,+\,\,\,\nabla{\vec{\sigma}}\,,\\
  \nonumber \\ 
  \label{eqn_lare_mhd_induction}
  \frac{\partial\vec{B}}{\partial t} & = & \nabla\times\Big(\,\vec{v}\times\vec{B}\,\Big)\,\,-\,\,\nabla\times\Bigg(\eta\frac{\nabla\times\vec{B}}{\mu_0}\Bigg)\,,\\
  \nonumber \\ 
  \label{eqn_lare_mhd_energy}
  \frac{\partial}{\partial t}\big(\,\rho\epsilon\big) & = & -\nabla\cdot(\,\rho\epsilon\vec{v}\,)\,\,\,-\,\,P\,\nabla\cdot\vec{v}\,\,+\,\,\eta J^{\,2}\,\,+\,\,\vec{\varepsilon}\,\vec{\sigma}\,,
\end{eqnarray}
with specific energy density,
\begin{eqnarray}
  \label{eqn_lare_mhd_epsilon}
  \epsilon & = & \frac{P}{(\gamma-1)\rho}\,,
\end{eqnarray}
where $\rho$ is the mass density, $\vec{v}$ is the plasma velocity, $\vec{B}$ the magnetic field, $P$ the thermal pressure, $\eta$ is the resistivity (not magnetic diffusivity), $J$ is the current density, $\gamma\,{=}\,5/3$ is the ratio of specific heats, and $\mu_0$ is the magnetic permeability. Viscous heating is represented by the last term of equation (\ref{eqn_lare_mhd_energy}), which also incorporates artificial viscosity (Wilkins 1980) in order to capture the heating effect of shocks. This heating term is expressed as the product of the rate of strain tensor,
\begin{eqnarray}
  \label{eqn_lare_strain_tensor}
  \varepsilon_{ij} & = & \frac{1}{2}\Bigg(\frac{\partial\,v_i}{\partial\,j}\,\,+\,\,\frac{\partial\,v_j}{\partial\,i}\Bigg)\,,
\end{eqnarray}
and the shock tensor,
\begin{eqnarray}
  \label{eqn_lare_shock_viscosity_tensor}
  \sigma_{ij} & = & \rho\,l\,\Big(\nu_1\,c_{\mathrm{f}}\,\,+\,\,\nu_2\,l\,|s|\Big)\,\Bigg(\varepsilon_{ij}\,\,-\,\,\frac{1}{3}\,\delta_{ij}\,\nabla\cdot\vec{v}\Bigg)\,,
\end{eqnarray}
where $c_{\mathrm{f}}$ is the fast magnetoacoustic speed, $l$ is the distance across a grid cell in the direction normal to the shock front, $s$ is a similarly localised strain rate (the subscripts $i$ and $j$ denote the different spatial coordinates), and the artificial viscosity coefficients $\nu_1\,{=}\,0.1$ and $\nu_2\,{=}\,0.5$ are constants. The form of equation (\ref{eqn_lare_shock_viscosity_tensor}) is derived from the Rankine-Hugoniot jump conditions and the values of the coefficients have been chosen such that numerical oscillations behind shock fronts are prevented. Note, the force equation (\ref{eqn_lare_mhd_force}) also acquires a viscous term. Gravitational effects are ignored in this study, as are thermal conduction and radiation. The simulations are concerned with how the magnetic field changes in response to the kink instability; specifically, how much magnetic energy is released and how the field subsequently evolves. Conduction becomes important some time after the energy release and later, radiation is the dominant process. Note, numerical studies have shown that conduction can act on MHD time scales (Botha et al. 2011): the amount of energy released from the field is unaffected, but the kinetic energy parallel to the field is much reduced.

The MHD equations are made dimensionless by replacing the variables with dimensionless equivalents. For example,
\begin{eqnarray}
  \nonumber r & = & \frac{r^*}{R_{\mathpzc{b}}}\,,\,\,\,\,\,\,\rho\,\,=\,\,\frac{\rho^*}{\rho_0}\,,\,\,\,\,\,\,B\,\,=\,\,\frac{B^*}{B_1}\,,
\end{eqnarray}
where asterisks denote dimensional variables, $R_{\mathpzc{b}}$ is the loop radius, $\rho_0$ the initial mass density, and $B_1$ the initial axial field at $r\,{=}\,0$. The other variables are expressed in a similar manner;
\begin{eqnarray}
  \nonumber \mathpzc{L} & = & \frac{\mathpzc{L}^*}{R_{\mathpzc{b}}}\,,\,\,\,\,\,\,t\,\,=\,\,\frac{t^*}{t_{\mathrm{A}}}\,,\,\,\,\,\,\,v\,\,=\,\,\frac{v^*}{v_{\mathrm{A}}}\,,\,\,\,\,\,\,P\,\,=\,\,\frac{P^*}{P_0}\,,
\end{eqnarray}
where $\mathpzc{L}^*\,{=}\,20\,R_{\mathpzc{b}}$ is the loop length, $t_{\mathrm{A}}\,{=}\,R_{\mathpzc{b}}/v_{\mathrm{A}}$ is the radial Alfv{\'e}n transit time, $v_{A}\,{=}\,B_1/\sqrt{\mu_{0}\rho_0}$ the Alfv{\'e}n speed, and $P_0\,{=}\,B_1^2/\mu_0$ the magnetic pressure. The specific energy density, current density and resistivity ($\epsilon$, $J$ and $\eta$) also have reference variables that can be expressed in terms of $R_{\mathpzc{b}}$, $\rho_0$ and $B_1$:
\begin{eqnarray}
  \nonumber \epsilon_0 & = & \frac{B_1^2}{\mu_0\rho_0}\,\,=\,\,v_{\mathrm{A}}^2\,,\,\,\,\,\,\,J_0\,\,=\,\,\frac{B_1}{\mu_0 R_{\mathpzc{b}}}\,,\,\,\,\,\,\,\eta_0\,\,=\,\,\mu_0 R_{\mathpzc{b}} v_{\mathrm{A}}\,.
\end{eqnarray}
Values appropriate for a coronal active region can be obtained by setting $R_{\mathpzc{b}}\,{=}\,1\,\mathrm{Mm}$, $\rho_0\,{=}\,1.6726\,{\times}\,10^{-13}\,\mathrm{kg}\,\mathrm{m}^{-3}$ and $B_1\,{=}\,50\,\mathrm{G}$. Hence, the length becomes $20\,\,\mathrm{Mm}$, $v_{A}\,{\approx}\,10\,\,\mathrm{Mm}\,\mathrm{s}^{-1}$ and $\eta_0\,{\approx}\,4\pi\,{\times}\,10^6\,\Omega\,\mathrm{m}$. 

The resistivity is taken to be non-uniform in these simulations,
\begin{eqnarray}
  \label{eqn_lare_mhd_resistivity}
  \nonumber \eta & = & \eta_{\mathrm{b}}\,,\hspace{1.95cm}|\,J\,| < J_{\mathrm{crit}}\,,\\
  \label{eqn_lare_mhd_resistivity_anomalous}
  \nonumber \eta & = & \eta_{\mathrm{b}}\,+\,\eta_{\mathrm{c}}\,,\hspace{1.15cm}|\,J\,| \geq J_{\mathrm{crit}}\,,
\end{eqnarray}
where $\eta_{\mathrm{b}}$ is the background resistivity (normally set to zero, since actual coronal resistivities are approximately $\mu_0\,\mathrm{m}^2\,\mathrm{s}^{-1}$) and $\eta_{\mathrm{c}}\,{=}\,0.001$ is the anomalous resistivity, which is only switched on when the current reaches or exceeds $J_{\mathrm{crit}}\,{=}\,15$. The value of $J_{\mathrm{crit}}$ is set so that it is significantly higher than the maximum current at the start of the simulation. Super-critical currents appear when, during the nonlinear evolution of the kink instability, current sheets begin to form and decrease in thickness. The anomalous resistivity is intended to capture the dissipation occurring at scales below the grid resolution: at this scale, resistivity is enhanced by small-scale plasma instabilities.

The computational domain is a 3D staggered grid: physical variables are not calculated at the same place for each cell in the domain. This approach improves numerical stability and allows conservation laws to be included in the computation. The domain size is $\mathpzc{L}_{x}\,{=}\,\mathpzc{L}_{y}\,{=}\,6$ (-3:+3) and $\mathpzc{L}_{z}\,{=}\,20$ (-10:+10). Initially, the loop axis follows the $z$-axis and the loop radius is $r\,{=}\,1$; therefore, the simulated loops all have an aspect ratio of 20. The loop is line-tied at $z\,{=}\,-10,\,+10$, which means, at those $z$-coordinates, the velocity components are set to zero. The velocity components are zero at the boundaries for all directions. The normal derivatives of magnetic field, energy and density are zero at all boundaries. The simulations are run with two grid resolutions: $128^2\,{\times}\,256$ (low) and $256^2\,{\times}\,512$ (high). It is assumed that a result is not a numerical artefact if it is consistent across both resolutions. 

\subsection{Initial configuration}
\label{sec_InitialConfiguration}
Some previous studies have used LARE3D to simulate the application of kink perturbations to a straightened line-tied coronal loop, see Gerrard et al. (2002), Gerrard \& Hood (2003), Browning et al. (2008) and Hood et al. (2009).
\begin{figure}[h!]  
  \center
  \includegraphics[scale=0.3]{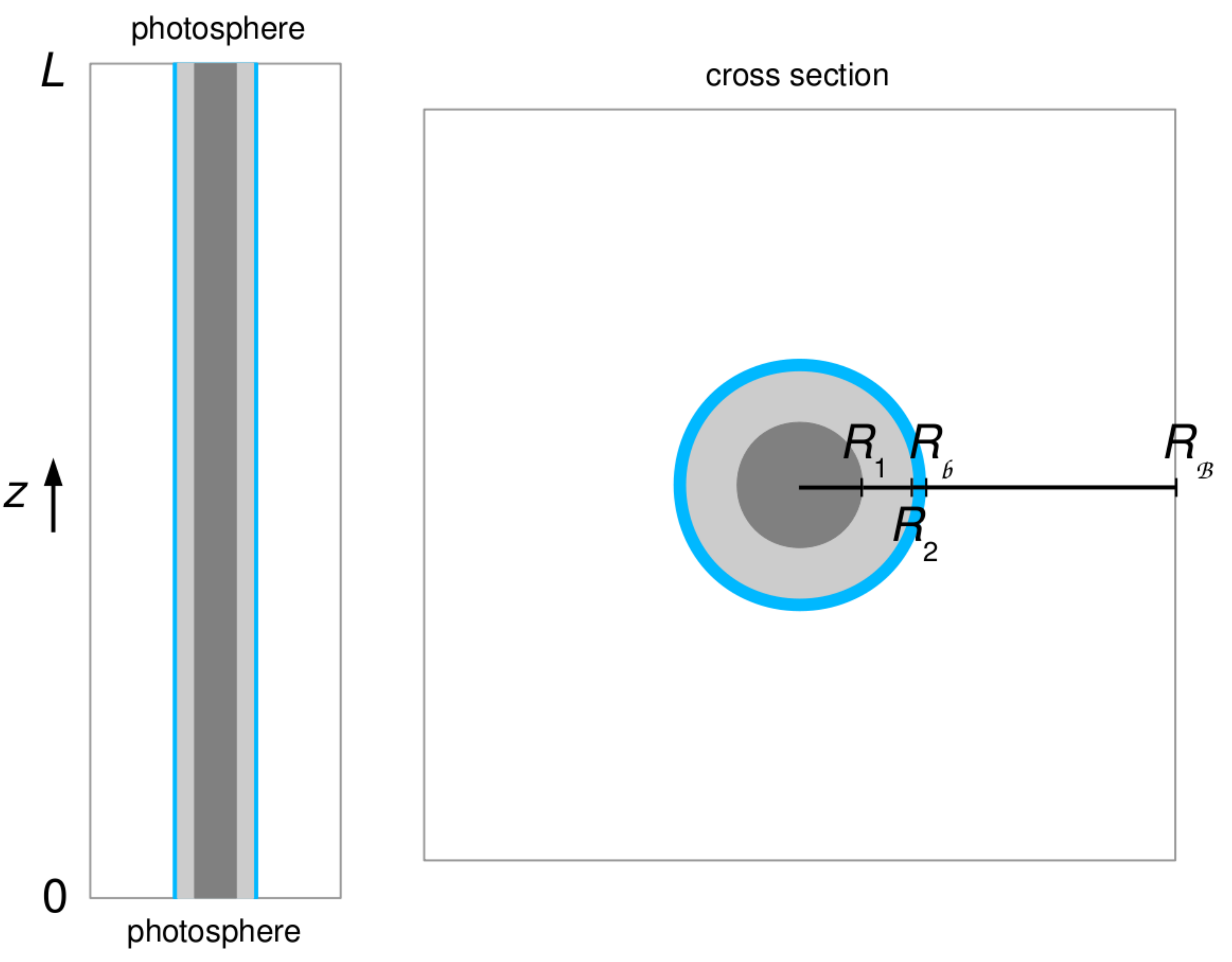}
  \caption{\small{Schematic of a straightened coronal loop in the $x\text{-}z$ plane (left) and in the $x\text{-}y$ plane (right). The loop, comprises a core (dark grey), an outer layer (light grey) and a current neutralisation layer (blue); the whole loop is embedded in a rectangular potential envelope. The core radius is half the loop radius ($R_1$:$R_2$:$R_{\mathpzc{b}}$:$R_{\mathpzc{B}}$ = 0.5:0.9:1:3, where $R_{\mathpzc{B}}$ is the distance from the initial loop axis to the edge of the envelope). The loop aspect ratio ($\mathpzc{L}/R_{\mathpzc{b}}$) in this figure is 20.}}
  \label{znc_schematic}  
\end{figure}
The initial equilibrium model used in the latter two studies was extended by Bareford et al. (2011) to include an outer current-neutralising layer so as to ensure the loop has (at least initially) zero net current: this improves on the model used by Browning \& Van der Linden (2003) and Bareford et al. (2010), which allowed loops to have net current (i.e., an azimuthal field was usually present in the potential envelope). All currents are now created by convective motions \textit{local} to the loop footpoints. Hence, a current neutralisation layer is introduced here, defined such that the azimuthal field ($B_{\theta}$) always falls to zero at the loop boundary ($R_{\mathpzc{b}}$); therefore, $B_{\theta}$ is zero in the potential envelope. 
The loop's radial $\alpha$-profile is approximated by a piecewise-constant function featuring three parameters (Figure \ref{znc_schematic}): the ratio of current to magnetic field is $\alpha_1$ in the core, $\alpha_2$ in the outer layer, $\alpha_3$ in the neutralisation layer and zero in the potential envelope. The free parameters are $\alpha_1$ and $\alpha_2$, whereas $\alpha_3$ is dependent on the first two and is determined by the requirement of zero net current. The magnetic field is continuous everywhere, whereas the current has discontinuities, and the outer surface of the potential envelope, representing the background corona, is placed at $R_{\mathpzc{B}}\,{=}\,3$ (three times the loop radius); this is far enough away that the boundary conditions do not influence the plasma evolution.

The fields are expressed in terms of the well-known Bessel function model, generalised to the concentric layer geometry (Melrose et al. 1994; Browning \& Van der Linden 2003; Browning et al. 2008). The field equations for the four regions (core, outer layer, neutralisation layer and envelope) are as follows:
\begin{eqnarray}
  \label{eqn_znc_field_equation_1}
  B_{1z} & = & B_{1}J_{0}(|\alpha_{1}|r)\,,\\
  \label{eqn_znc_field_equation_2}
  B_{1\theta} & = & \sigma_{1}B_{1}J_{1}(|\alpha_{1}|r)\,, \mbox{\hspace{3.1cm}} 0 \leq r \leq R_{1}\,,\\
  \nonumber &  &\\  
  \label{eqn_znc_field_equation_3}
  B_{2z} & = & B_{2}J_{0}(|\alpha_{2}|r) + C_{2}Y_{0}(|\alpha_{2}|r)\,,\\
  \label{eqn_znc_field_equation_4}
  B_{2\theta} & = & \sigma_{2}(B_{2}J_{1}(|\alpha_{2}|r) + C_{2}Y_{1}(|\alpha_{2}|r))\,, \mbox{\hspace{0.7cm}} R_{1} \leq r \leq R_{2}\,,\\
  \label{eqn_znc_field_equation_5}
  \nonumber &  &\\  
  B_{3z} & = & B_{3}J_{0}(|\alpha_{3}|r) + C_{3}Y_{0}(|\alpha_{3}|r)\,,\\
  \label{eqn_znc_field_equation_6}
  B_{3\theta} & = & \sigma_{3}(B_{3}J_{1}(|\alpha_{3}|r) + C_{3}Y_{1}(|\alpha_{3}|r))\,, \mbox{\hspace{0.7cm}} R_{2} \leq r \leq R_{\mathpzc{b}}\,,\\
  \nonumber &  &\\  
  \label{eqn_znc_field_equation_7}
  B_{4z} & = & B_{4}\,,\\
  \label{eqn_znc_field_equation_8}
  B_{4\theta} & = & 0\,, \mbox{\hspace{4.65cm}} R_{\mathpzc{b}} \leq r \leq R_{\mathpzc{B}}\,,  
\end{eqnarray}
where $\sigma_i\,=\,\frac{\alpha_i}{|\alpha_i|}$ ($i\,{=}\,1,2,3$) represents the sign of $\alpha_i$. The fields must be continuous at the inner radial boundaries, $R_1$, $R_2$ and $R_{\mathpzc{b}}$. (The positions are $R_1\,{=}\,0.5$, $R_2\,{=}\,0.9$ and $R_{\mathpzc{b}}\,{=}\,1$, so that most of the loop is similar to the one described by Bareford et al. (2010), but with a thin current neutralisation layer between $R_2$ and $R_{\mathpzc{b}}$.) Therefore, the coefficients $B_\textit{j}$ and $C_\textit{j}$ ($j\,{=}\,2,3,4$) are determined by the requirement of continuity of the magnetic field at all interfaces (Bareford et al. 2011). The value of $\alpha_3$ (the neutralisation layer current) is found, for a given ($\alpha_1{,}\,\alpha_2$), by numerical solution of $B_{3\theta}(R_{\mathpzc{b}})\,{=}\,0$, ensuring that the net current is zero and that the azimuthal field vanishes outside the loop, see Eq. (\ref{eqn_znc_field_equation_6}). From the nondimensionalisation of the magnetic field, the field coefficient at the core is $B_1\,{=}\,1$.

\begin{figure*}
  \center
  \vspace{10pt}
  \includegraphics[scale=0.45]{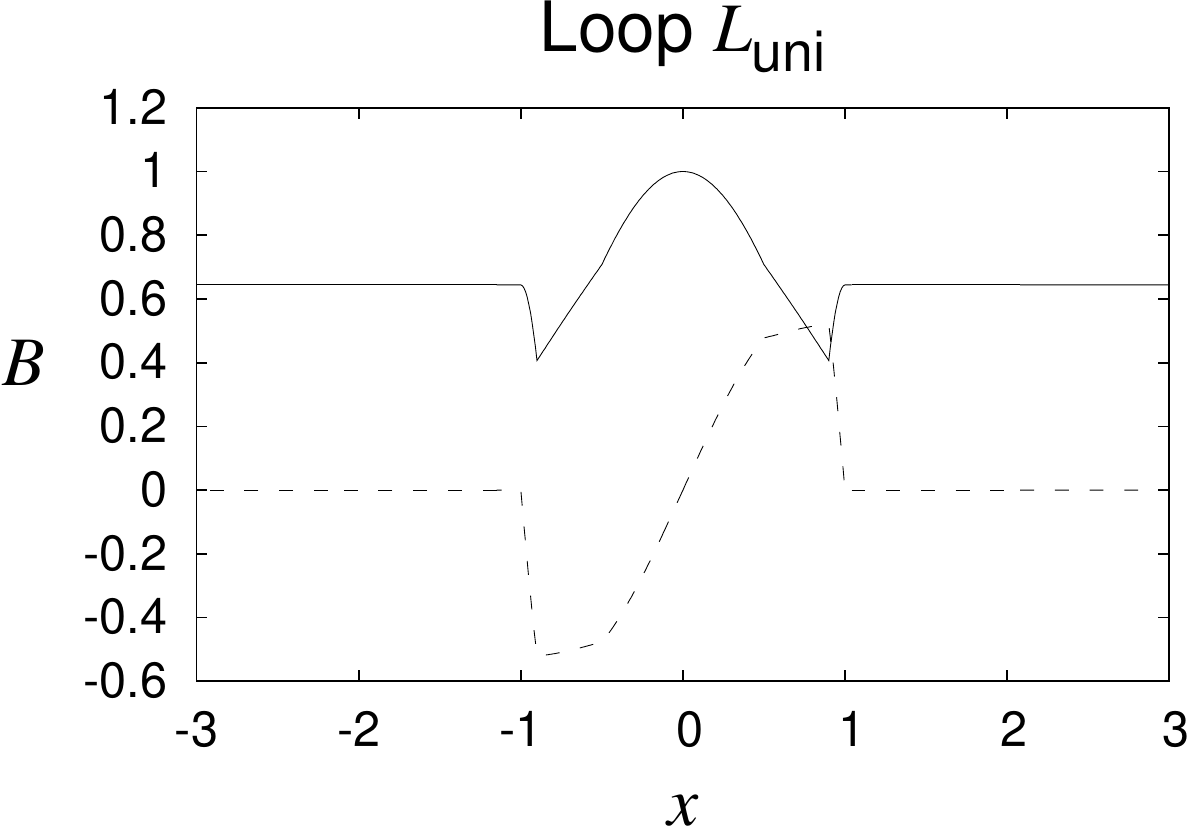}\hspace{10pt}
  \includegraphics[scale=0.45]{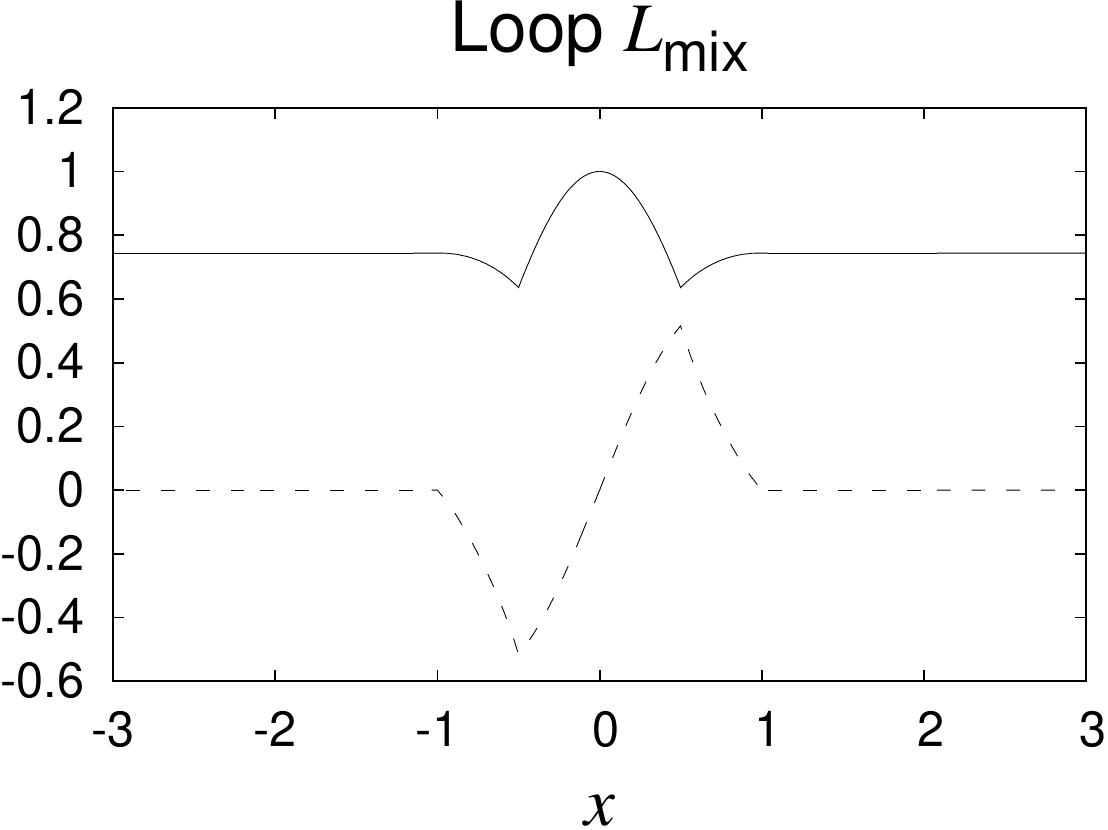}\hspace{10pt}
  \includegraphics[scale=0.45]{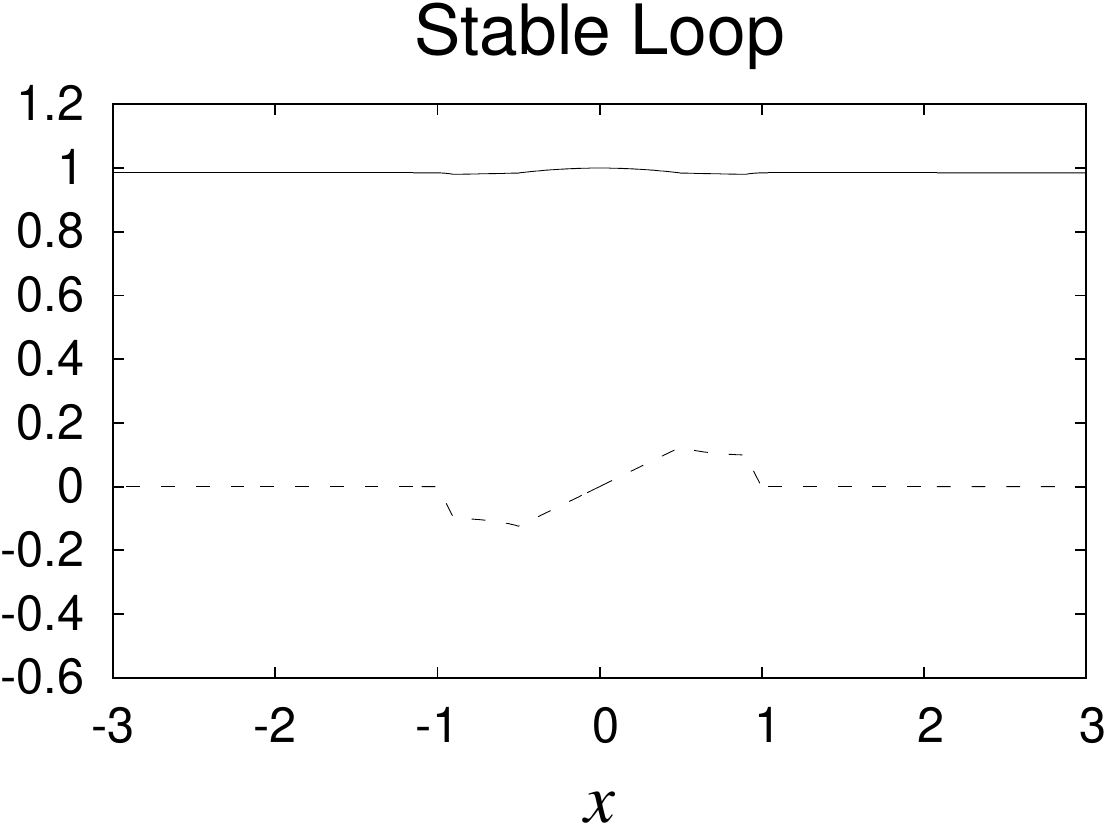}
  \caption{\small{The initial analytically-determined $B_{z}$ (solid) and $B_{\theta}$ (dashed) profiles for $L_{\mathrm{uni}}$ and $L_{\mathrm{mix}}$. The field profiles for the stable loop (last row of Table \ref{tab_loop_configs}) are also shown.}}
  \label{znc_b_rx_pf}  
\end{figure*}

The linear kink instability threshold for this current-neutralised loop was determined by the CILTS code (Van der Linden 1991; Browning \& Van der Linden 2003) --- it uses a bicubic Hermite finite element method to calculate the growth rates and eigenfunctions for specific line-tied $\alpha$-configurations.
\begin{figure}[h!]
  \center  
  \includegraphics[scale=0.7]{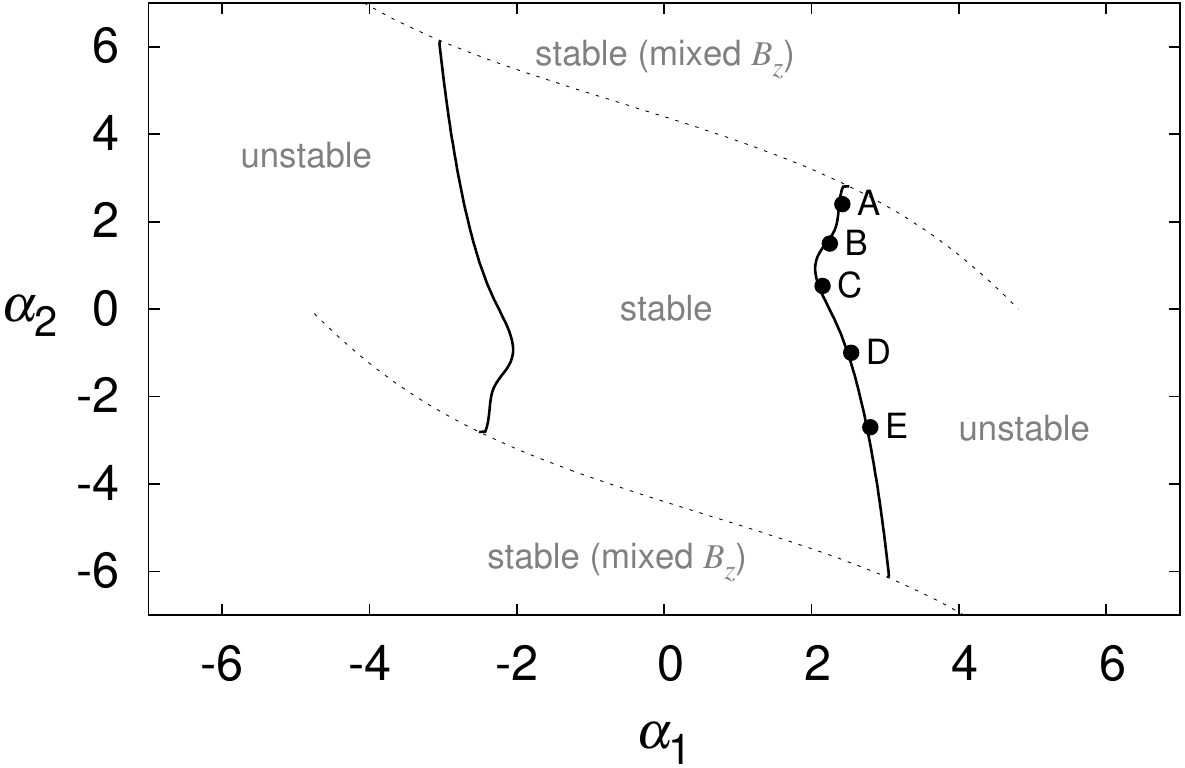}
  \caption{\small{The linear kink instability thresholds for $\mathpzc{L}/R_{\mathpzc{b}}\,{=}\,20$. These thresholds have been cropped by a pair of dashed lines that indicate where $B_z(r)$ starts to acquire a mixed polarity. The right threshold is sampled by a selection of five marginally unstable configurations (black circles).}}
  \label{znc_it_a2p_larepts}
\end{figure}
In contrast to the stability space for a loop of net current (Bareford et al. 2010), the instability threshold is open, see Figure \ref{znc_it_a2p_larepts}. This is very similar to the threshold found for a close-fitting conducting shell (Browning \& Van der Linden 2003), since in the case of zero net current, perturbations quickly fall to zero beyond the loop radius. A consequence of the Bessel function model is that the axial field changes sign if the values of $\alpha_1$ and $\alpha_2$ are too large. This reversal is unphysical and hence, introduces a restriction to the stable region of Figure \ref{znc_it_a2p_larepts}; namely, all stable configurations should have $B_{z}(r)$ of uniform sign. The new stability space is closed by excluding those field profiles that have a $B_z(r)$ of mixed polarity, since this could not be achieved directly by footpoint motions of an initially unidirectional field. The filled circles of Figure \ref{znc_it_a2p_larepts} identify the loop configurations (see also Table \ref{tab_loop_configs}) that will be simulated by the LARE3D code (the initial field profiles for Loops B and D are given in Figure \ref{znc_b_rx_pf}). All of these configurations are unstable to the ideal kink instability.

Loops of uniform twist (loops A--C) are a more likely result of \textit{correlated} convective driving (Bareford et al. 2011), where the threshold is approached via a series of steps with $\delta\alpha_1\,{\approx}\,\delta\alpha_2$ --- this is the reason why the part of the threshold curve where $\alpha_2\,{>}\,0$ is more finely sampled compared to $\alpha_2\,{<}\,0$. Note, the section of threshold on the left of Figure \ref{znc_it_a2p_larepts} is merely the negative of the section on the right; hence, it is not necessary to sample both threshold sections.
\begin{center}
  \begin{table*}
    \caption{The $\alpha$-profiles, axial fluxes and field coefficients for the simulated loops.}
    \label{tab_loop_configs}
    \begin{center}    
    \begin{tabular}{ l  c  c  c  c  c  c  c  c  c }
      \hline      
      \textbf{Loop} & \textbf{$\alpha_1$} & \textbf{$\alpha_2$} & \textbf{$\alpha_3$} & \textbf{$\psi_{\mathpzc{B}}$} & \textbf{$B_2$} & \textbf{$C_2$} & \textbf{$B_3$} & \textbf{$C_3$} & \textbf{$B_4$} \\ \hline     
      \textbf{A} & 2.42 & 2.4 & -13.08 & 15.3 & 0.1 & -0.0079 & 2.42 & -0.59 & 0.55 \\ \hline
      \textbf{B ($L_{\mathrm{uni}}$)} & 2.25 & 1.5 & -8.71 & 18.1 & 0.79 & -0.2 & -0.27 & 2.37 & 0.64 \\ \hline
      \textbf{C} & 2.15 & 0.53 & -4.95 & 20.8 & 0.61 & -0.15 & -0.94 & -1.85 & 0.74 \\ \hline
      \textbf{D ($L_{\mathrm{mix}}$)} & 2.54 & -1.0 & -0.84 & 21.0 & 0.91 & 0.50 & 0.92 & 0.38 & 0.74 \\ \hline
      \textbf{E} & 2.8 & -2.7 & 3.82 & 20.4 & 0.26 & 1.32 & -1.79 & 0.026 & 0.72 \\ \hline
      \textbf{Stable} & 0.5 & 0.1 & -0.96 & 27.8 & 0.97 & -0.0078 & 1.21 & 0.63 & 0.98 \\
      \hline    
    \end{tabular}
    \end{center}
  \end{table*}
\end{center}
The configurations listed in Table \ref{tab_loop_configs} encompass two types of loop; one class where $\alpha_1$ and $\alpha_2$ have the same sign (A--C) and the other where these parameters have opposing signs \text{(D and E)}. Loop B has been chosen as representative of the first loop type and Loop D of the second. Henceforth, Loop B will be referred to as $L_{\mathrm{uni}}$ (uniform twist) and Loop D will be denoted by $L_{\mathrm{mix}}$ (mixed twist).

\begin{sloppypar}
Each of the loops indicated in Figure \ref{znc_it_a2p_larepts} is subjected to a disturbance of the form,
\begin{eqnarray}
  \label{eqn_perturbed_radial_velocity}
  v_r & = & c_1 e^{-4 r^4}\Bigg[\cos{\Bigg(\frac{\pi}{\mathpzc{L}} z\Bigg)}\cos\big(kz-\theta\big)\Bigg]\,,
\end{eqnarray}
where $v_r$ is the radial component of the perturbed velocity, $\mathpzc{L}$ is the loop length, $r\,{=}\,\sqrt{x^2+y^2}$ is the radial coordinate, $k\,{=}\,1.1$ is the wave number, $\theta\,{=}\,\arctan(x/y)$, and the constant $c_1\,{=}\,0.01$ reduces the amplitude so that the perturbation is initially linear. This disturbance initiates the kink instability.
\end{sloppypar}
Both $L_{\mathrm{uni}}$ and $L_{\mathrm{mix}}$ are simulated for low and high resolutions; only the high resolution is used for the other loop configurations. Each simulation runs until the magnetic field appears to have settled into a lower energy state.

\section{Numerical results}
\label{sec_NumericalResults}

\subsection{Energy and resistivity}
\label{sec_NumericalResults_EnergyResistivity}
\begin{figure*}
  \center
  \includegraphics[scale=0.45]{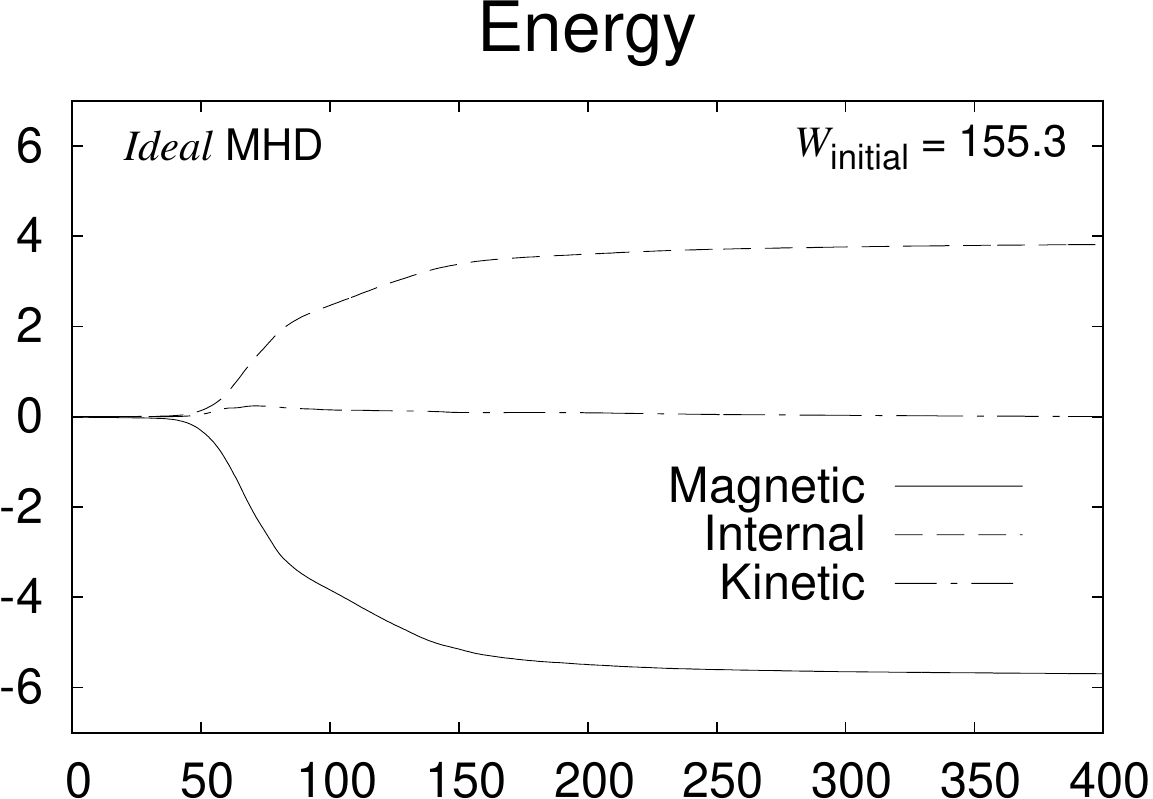}\hspace{5pt}
  \includegraphics[scale=0.45]{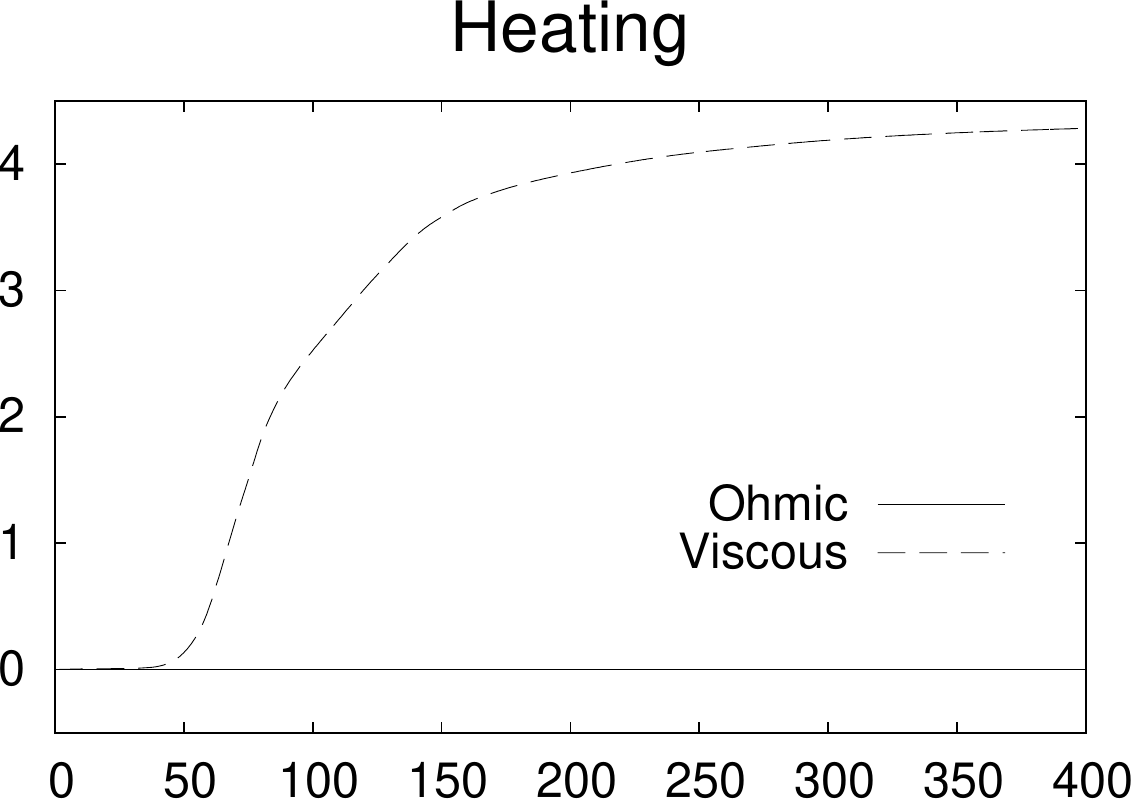}\hspace{5pt}
  \includegraphics[scale=0.45]{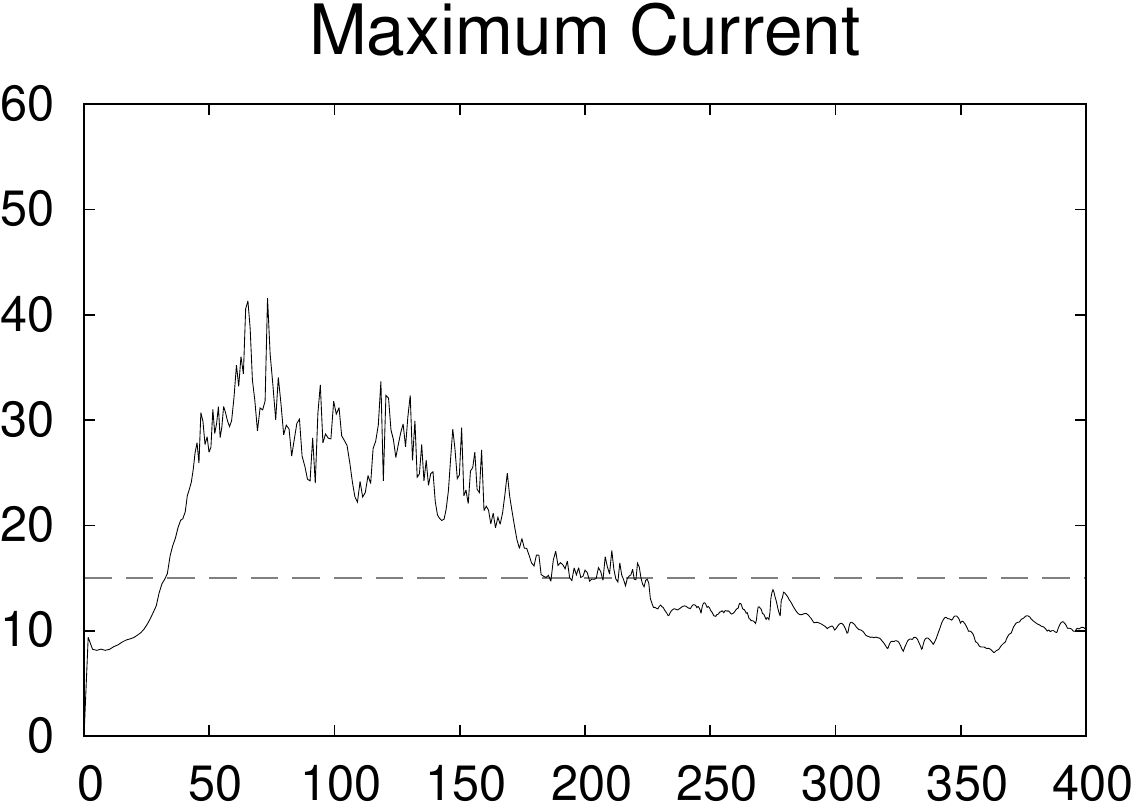}\vspace{10pt}
  \includegraphics[width=150pt,height=89pt]{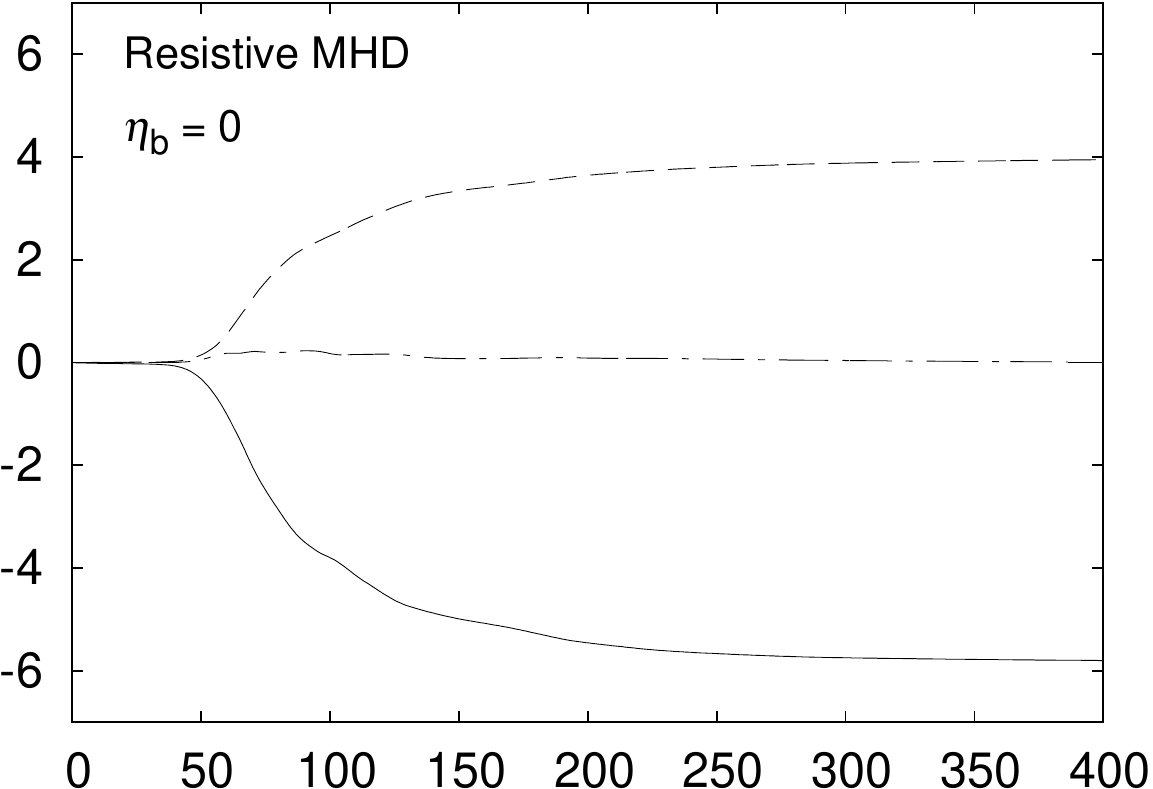}\hspace{4pt}
  \includegraphics[width=150pt,height=89pt]{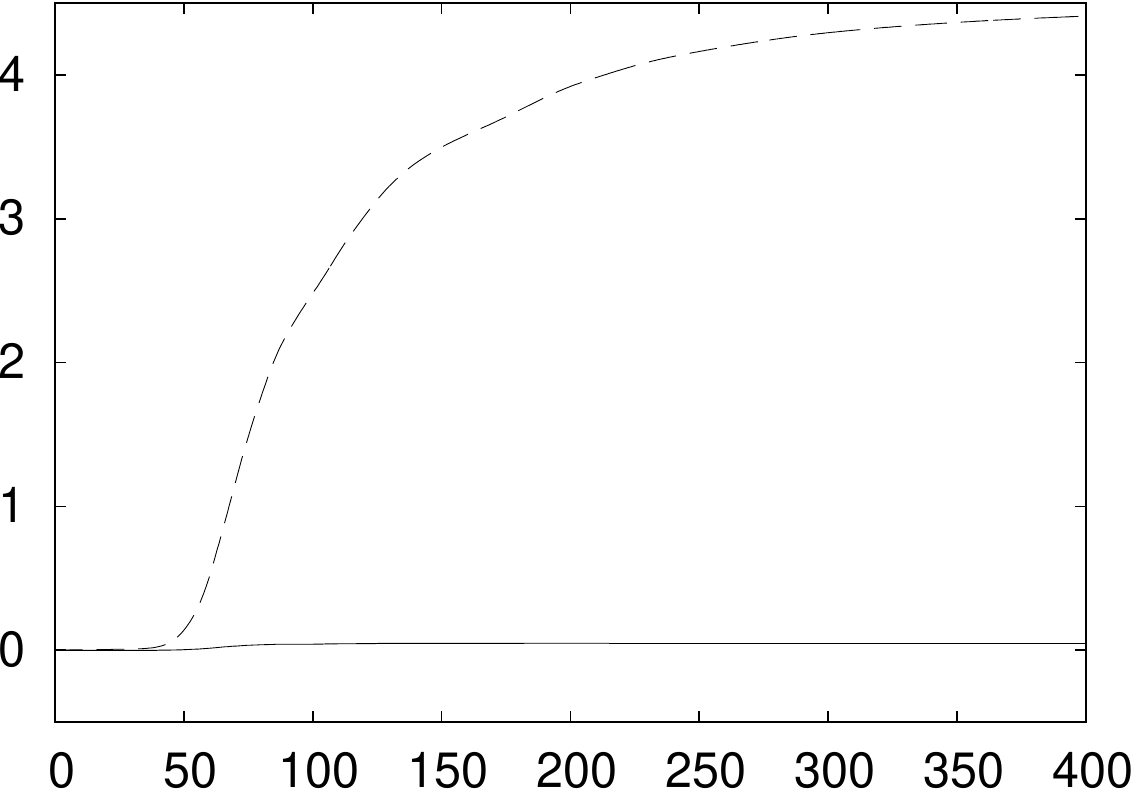}\hspace{4pt}
  \includegraphics[width=148pt,height=91pt]{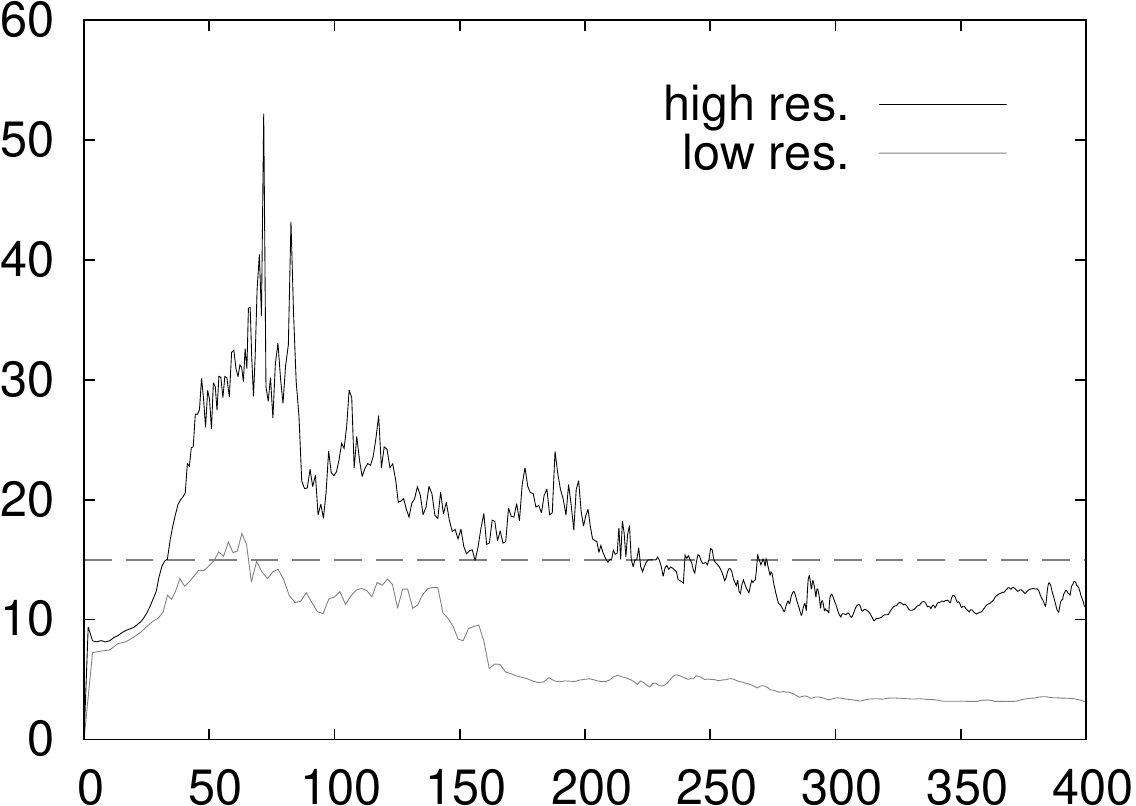}\vspace{10pt}
  \includegraphics[scale=0.45]{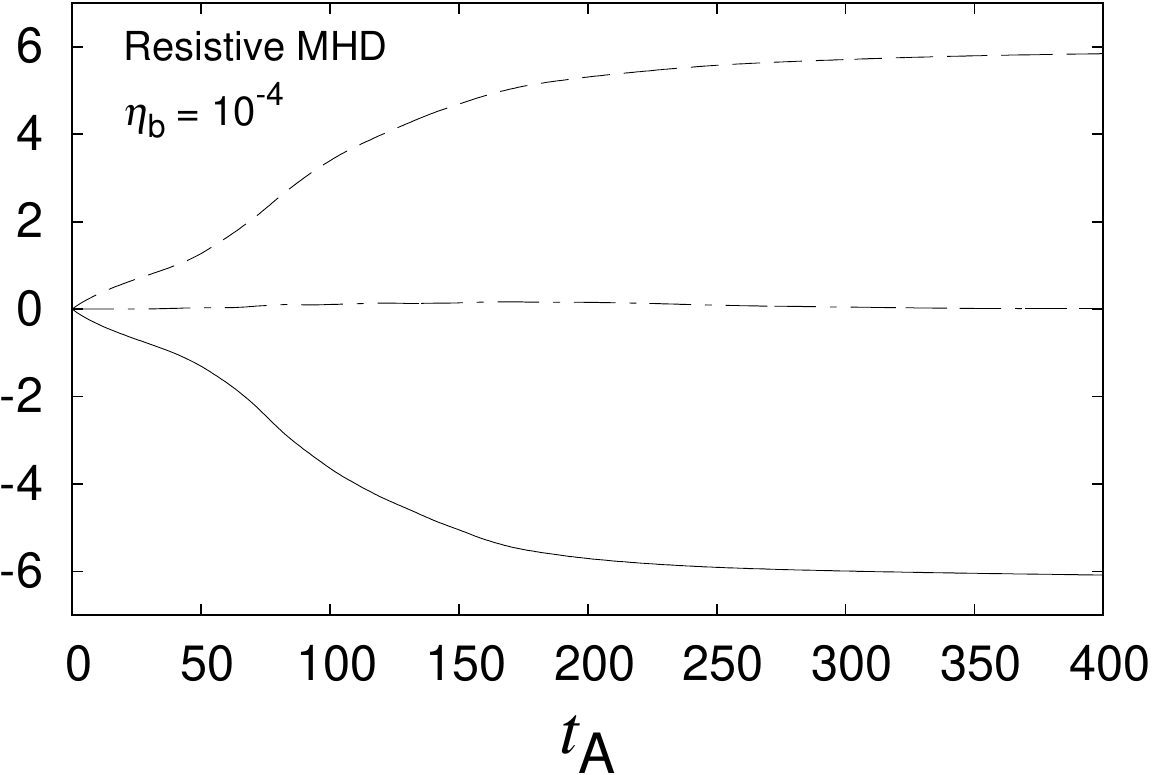}\hspace{5pt}
  \includegraphics[scale=0.45]{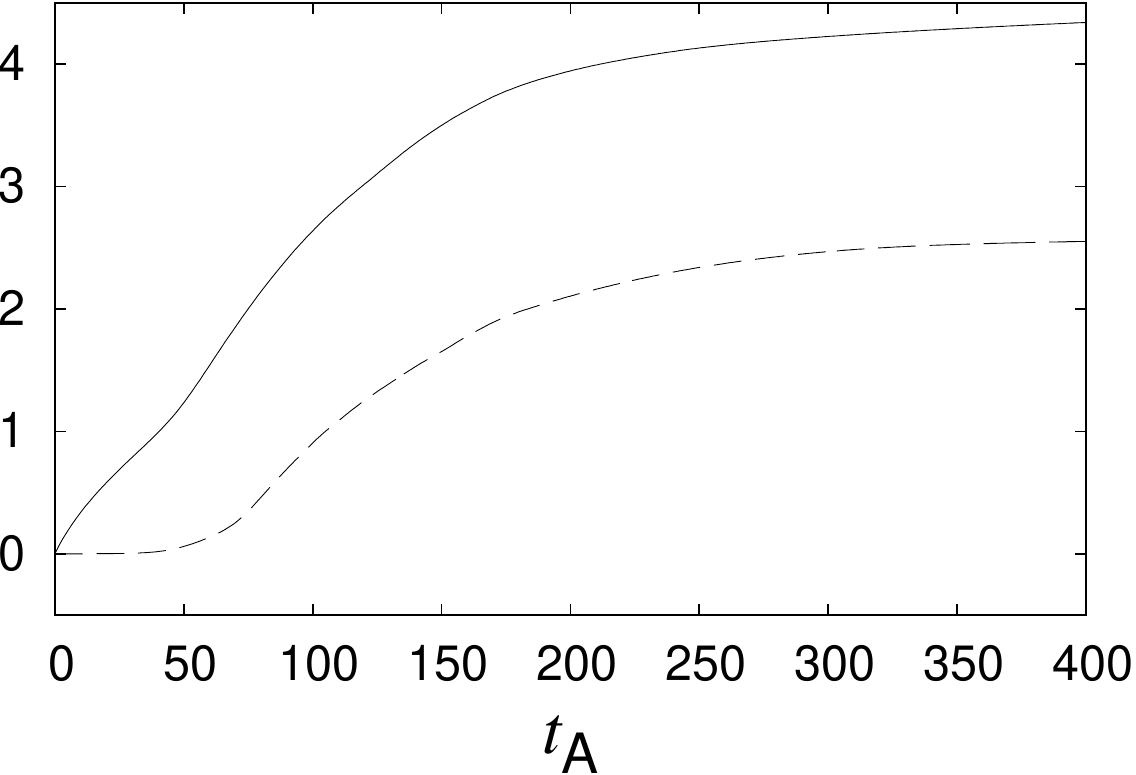}\hspace{5pt}
  \includegraphics[scale=0.45]{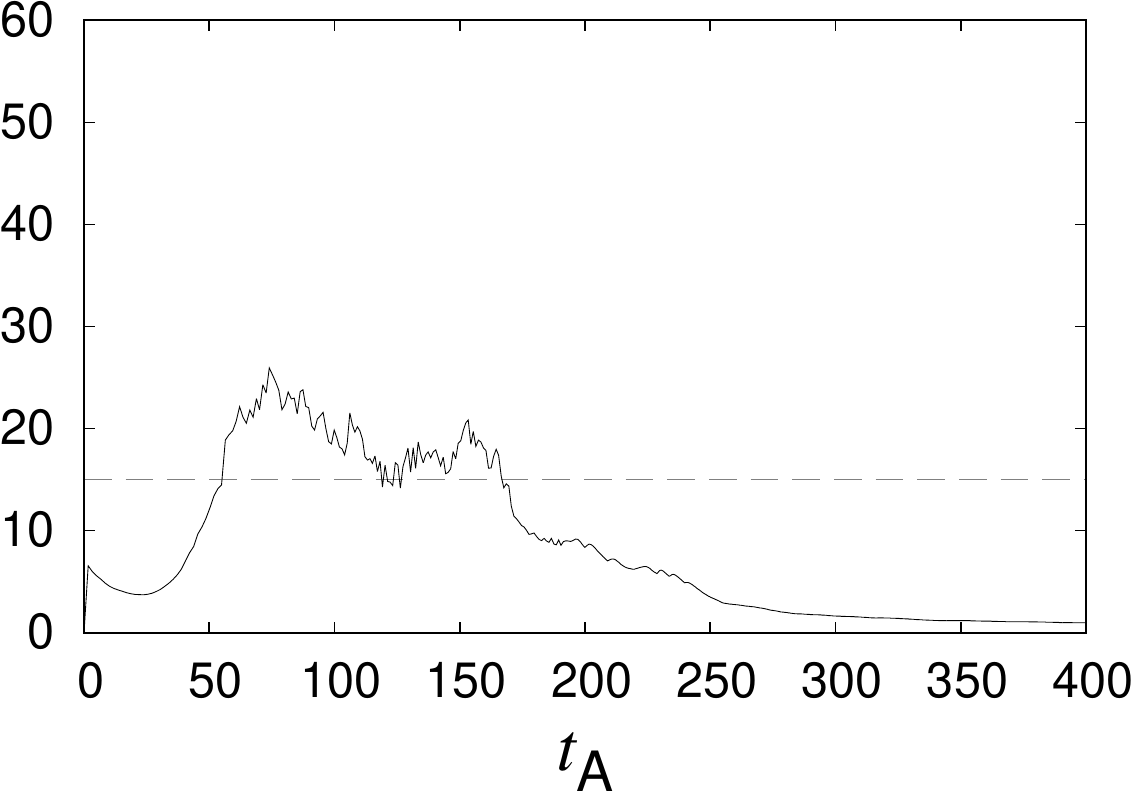}
  \caption{\small{Loop $L_{\mathrm{uni}}$: the temporal variation in energy (magnetic, internal and kinetic), heating (Ohmic and viscous) and maximum current for \textit{ideal} MHD (top row), for resistive MHD with $\eta_{\mathrm{b}}\,{=}\,0$ (middle row), and for $\eta_{\mathrm{b}}\,{=}\,10^{-4}$ (bottom row). The initial magnetic energy has been subtracted from the magnetic energy plots (solid lines, left column). The critical current ($J_{\mathrm{crit}}\,{=}\,15$) is indicated by the grey dashed horizontal lines (right column). For the $\eta_{\mathrm{b}}\,{=}\,0$ case, the maximum current plots are from the high (black) and low (grey) resolution simulations.}}
  \label{b_energy_heating_jmax_plots}
\end{figure*}

Loop $L_{\mathrm{uni}}$ ($\alpha_1\,{=}\,2.25$, $\alpha_2\,{=}\,1.5$) is linearly kink unstable, and the numerical simulation (Figure \ref{b_energy_heating_jmax_plots}, middle row) shows that it is also \textit{nonlinearly} unstable. The magnetic energy, $W$, has an initial (dimensionless) value of 155.3 when integrated over the entire simulation domain,
\begin{eqnarray}
  \nonumber W & = & \frac{1}{2\mu_0} \int B^2\,\,dV\,;
\end{eqnarray}
the internal ($U$) and kinetic ($E$) energies are also volume integrals,
\begin{eqnarray}
  \nonumber U & = & \frac{1}{\gamma-1} \int P\,\,dV\,,\,\,\,\,\,\,E\,=\,\,\frac{1}{2} \int \rho v^2\,\,dV\,.
\end{eqnarray}
The initial magnetic energy integrated over the loop $L_{\mathrm{uni}}$ only is much less ($W_{\mathpzc{b}}\,{\approx}\,20$). All magnetic energies can be dimensionalised by making a simple correction to equation (27) given in Bareford et al. (2010): since here the axial flux is \textit{not} normalised to 1, the dimensionalising multiplier must first be divided by $\psi_{\mathpzc{B}}^2$, which is the square of the non-dimensional axial flux over the radius $R_{\mathpzc{B}}$ (Table \ref{tab_loop_configs}, column 5). For loop $L_{\mathrm{uni}}$, this gives a multiplier of $4.8\,{\times}\,10^{19}$ (assuming a radius of $1\,\mathrm{Mm}$ and a mean axial field strength of $50\,\mathrm{G}$): thus, the dimensionalised initial loop energy is roughly $10^{21}\,\mathrm{J}$.

At the onset of instability, $W$ undergoes a decrease, coincident with a rise in $U$ and with a much more modest increase in kinetic energy ($E_{\mathrm{max}}\,{\approx}\,0.2$). The maximum current, $J_{\mathrm{max}}$, also rises just before the decrease in $W$, indicating the formation of a helical current sheet. The nonlinear instability starts at around $t\,{=}\,50\,t_{\mathrm{A}}$ and within the next $50\,t_{\mathrm{A}}$, approximately 70\% of the total energy release has been achieved. Magnetic energy reduces more slowly after $t\,{=}\,100\,t_{\mathrm{A}}$. The release of magnetic energy is of a similar size for both resolutions, and significantly, larger currents are recorded at high resolution, which is expected; otherwise, spatially-confined changes in current are missed and anomalous resistivity is reduced. The fact that the maximum current increases with resolution is indicative of current sheet formation. These structures have (possibly) infinite current density, so higher resolutions should reveal larger values of the maximum.

In the top two cases of Figure \ref{b_energy_heating_jmax_plots}, \textit{ideal} MHD and zero background resisitivity, it can be seen that there is no detectable change in energy until around $40\,t_{\mathrm{A}}$ --- this is the time when a current sheet starts to form. The adjective \textit{ideal} is italicised because the rate of magnetic diffusion in the corona is easily exceeded by numerical resistivity; therefore, any reduction in magnetic energy associated with a \textit{stable} configuration is artificial. Hence, for the stable case, in which no current sheets ever form, there is no sudden onset of dissipation, only gradual dissipation as expected. In the bottom row of Figure \ref{b_energy_heating_jmax_plots} ($\eta_{\mathrm{b}}\,{=}\,10^{-4}$), there is a continual dissipation of magnetic energy due to Ohmic effects; however, even here, there is a clear onset of enhanced dissipation at the point of current sheet formation.
This all agrees with previous work (Browning et al. 2008; Hood et al. 2009) and clearly indicates that energy dissipation (conversion from magnetic energy to internal energy) is closely associated with the existence of current sheets.

However, Figure \ref{b_energy_heating_jmax_plots} also shows that energy is not conserved for the two cases mentioned (\textit{ideal} MHD and $\eta_{\mathrm{b}}\,{=}\,0$). At high resolution, only ${\sim}\,68\%$ of the energy released from the magnetic field is converted to internal energy, and by the end of the simulation, less than one percent of the energy release is in the form of kinetic energy. Halving the spatial resolution worsens the energy conservation by almost $10\%$. This loss of energy is due to spatially unresolved current sheets, resulting in numerical diffusion. The results for \textit{Ideal} MHD (i.e., zero magnetic diffusivity) are very similar to the ones produced for the resistive MHD case with no background resistivity. This shows that, for these cases, Ohmic dissipation does \textit{not} contribute significantly to the rise in internal energy, which is instead mainly caused by viscous heating. The use of a non-zero background resistivity ($\eta_{\mathrm{b}}\,{=}\,10^{-4}$) changes the plots. The reduction in magnetic energy starts right away and is more drawn out and the maximum current is much less than when $\eta_{\mathrm{b}}\,{=}\,0$. These differences are all consistent with an increased resistivity. Crucially, Ohmic heating is now clearly present and, with the viscous heating, accounts for nearly all of the dissipated magnetic energy. In terms of energy conservation, LARE3D performs better when $\eta_{\mathrm{b}}\,{>}\,0$: the internal energy increase is now approximately $96\%$ of the magnetic energy decrease (the final kinetic energy is less than $0.2\%$ of $\delta{W}$). A background resistivity of $10^{-4}$ appears to be the smallest value that effectively minimises the effect of the numerical resistivity, such that the results are likely to be a reasonable description of the thermodynamics. If $\eta_{\mathrm{b}}$ is halved, the percentage of the magnetic energy release lost to the simulation increases to around $10\%$. Further tests have shown that energy conservation is not improved by a lowering the critical current and thereby causing the anomalous resistivity ($\eta_{\mathrm{c}}$) to be applied earlier in the simulation.

A sufficiently large background resistivity does substantially mitigate the amount of energy lost by numerical resistivity, at least for $L_{\mathrm{uni}}$; but $\eta_{\mathrm{b}}\,{=}\,10^{-4}$ is not realistic by coronal standards, and Figure \ref{b_energy_heating_jmax_plots} (bottom left) reveals a significant drop in field energy before the kink instability takes effect. This initial decline is caused by global Ohmic diffusion rather than magnetic reconnection, since current sheets have not yet formed. The numerical resistivity comes about when current sheet widths begin to fall below the grid resolution; although magnetic energy continues to be dissipated, it is not converted to other forms of energy (e.g., internal or kinetic). However, shocks can still be resolved during the unstable phase, and these contribute to the internal energy via viscous heating. If the background resistivity is large enough it will limit currents and thereby prevent current sheets from becoming too thin. Numerical (that is to say artificial) resistivity will be a factor during the conversion process when $\eta_{\mathrm{b}}\,{=}\,0$: approximately $32\%$ of the reduction in magnetic field energy is not accounted for by the final internal energy --- this is also true for the other loop configurations.
\begin{figure*}
  \center
  \includegraphics[width=150pt,height=115pt]{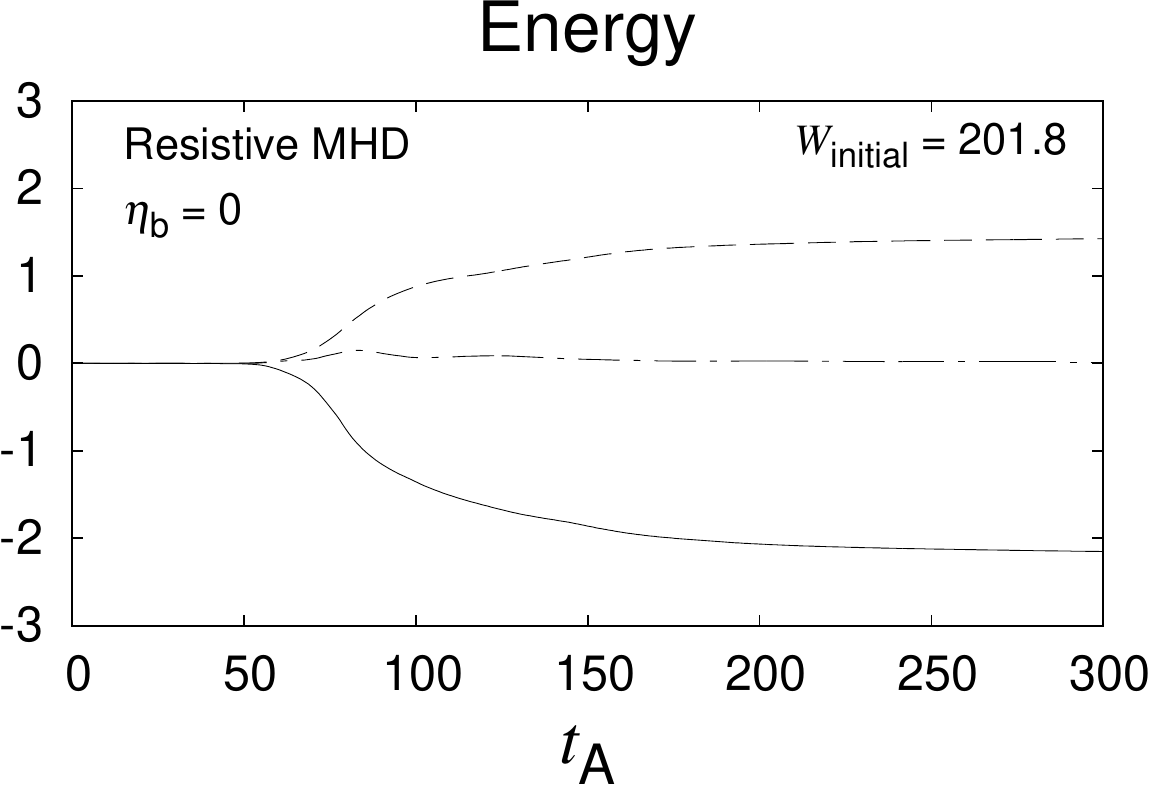}\hspace{4pt}
  \includegraphics[width=150pt,height=115pt]{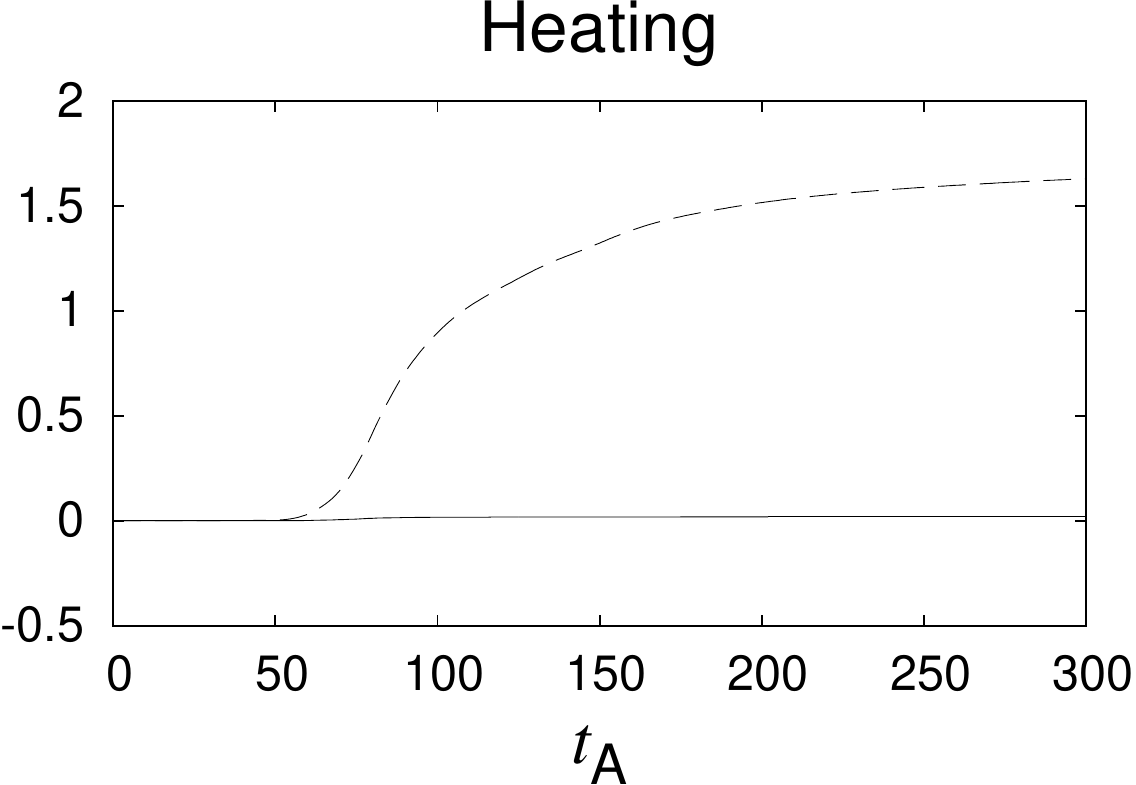}\hspace{4pt}
  \includegraphics[width=150pt,height=115pt]{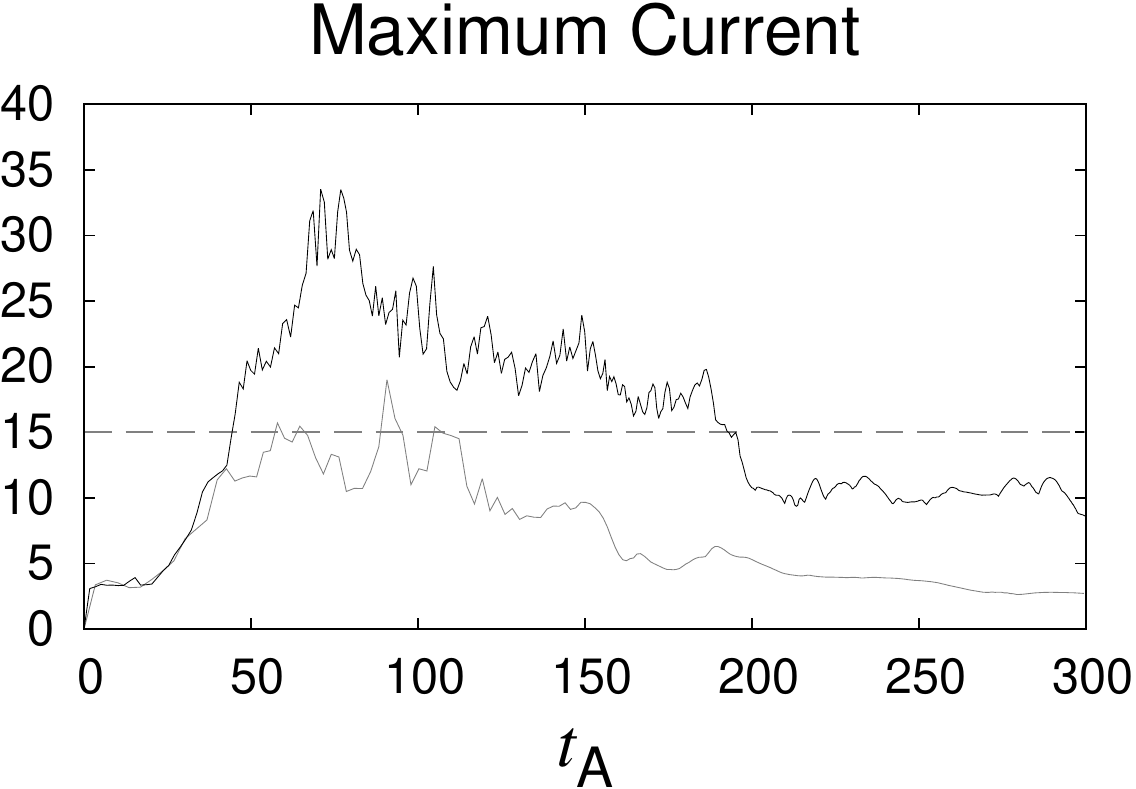}
  \caption{\small{Loop $L_{\mathrm{mix}}$: the temporal variation in energy (magnetic, internal and kinetic), heating (Ohmic and viscous) and maximum current (low and high resolution) for resistive MHD with $\eta_{\mathrm{b}}\,{=}\,0$. The different plot lines follow the same scheme as that used for Figure \ref{b_energy_heating_jmax_plots}.}}
  \label{d_energy_heating_jmax_plots}
\end{figure*}
However, virtually all of this artificial resistivity is occurring during the energy conversion. The drop in magnetic energy is a robust result. To demonstrate further, we have used \textit{ideal} MHD to simulate a loop ($\alpha_1\,{=}\,0.5$ and $\alpha_2\,{=}\,0.1$, see Table \ref{tab_loop_configs}, bottom row) that is well within the stable region (as shown in Figure \ref{znc_it_a2p_larepts}). In the absence of an instability (and any applied resistivity), the energy declines by around one thousandth of one percent over $300\,t_{\mathrm{A}}$; this reduction is three orders of magnitude smaller than the energy release caused by the kink instability.

Loop $L_{\mathrm{mix}}$ ($\alpha_1\,{=}\,2.54$, $\alpha_2\,{=}\,{-}1.0$) has a core that is oppositely twisted with respect to its outer layer. Figure \ref{d_energy_heating_jmax_plots} shows the same correspondence between the magnetic and internal energies that was seen for the previous loop. Again, the magnetic energy release is of the same size for both resolutions and, at the higher resolution, significantly larger currents are recorded; although the size of the release is around half that found for $L_{\mathrm{uni}}$. Note, the initial magnetic energy is higher than the value given for the other loop. This is a consequence of setting $B_1\,{=}\,1$: $L_{\mathrm{mix}}$ is less twisted and therefore has a higher axial flux, which results in a lower \textit{dimensionalised} magnetic energy.

The general trends for magnetic, internal and kinetic energies are consistent between resolutions (for both loops), and most importantly, so is the size of the magnetic energy release. Therefore, simulations at higher resolution are not required --- this paper will proceed with results taken only from the $256^2\,{\times}\,512$ simulations. The results for the other loop configurations on the threshold (Figure \ref{znc_it_a2p_larepts}) also suggest that linear instability evolves to a nonlinear stage, which gives rise to current sheets, magnetic reconnection and most importantly, shock heating. The following sections will present results based on a current-dependent resistivity and zero background resistivity --- this is more compatible with the coronal environment. The breakdown of energy conservation associated with this parameter choice can be ignored since we are only interested in how the magnetic field behaves during and after the instability. However, this issue will have to be addressed for detailed studies of the thermodynamic evolution.

\subsection{Magnetic field and current magnitude}
\label{sec_NumericalResults_MagneticField_CurrentMagnetic}
\begin{figure*}
  \center
  \includegraphics[scale=0.11]{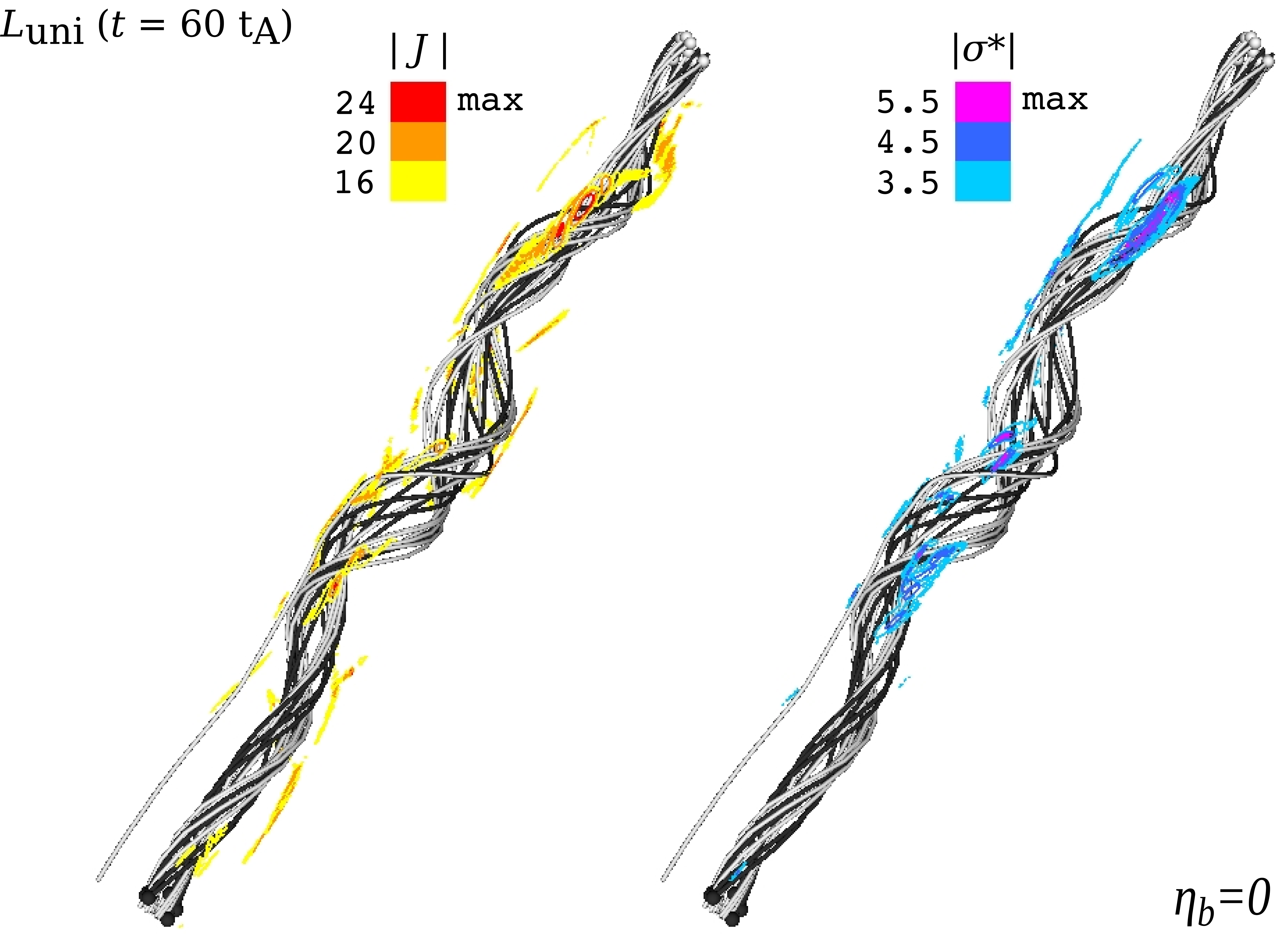}\vspace{10pt}
  \includegraphics[scale=0.11]{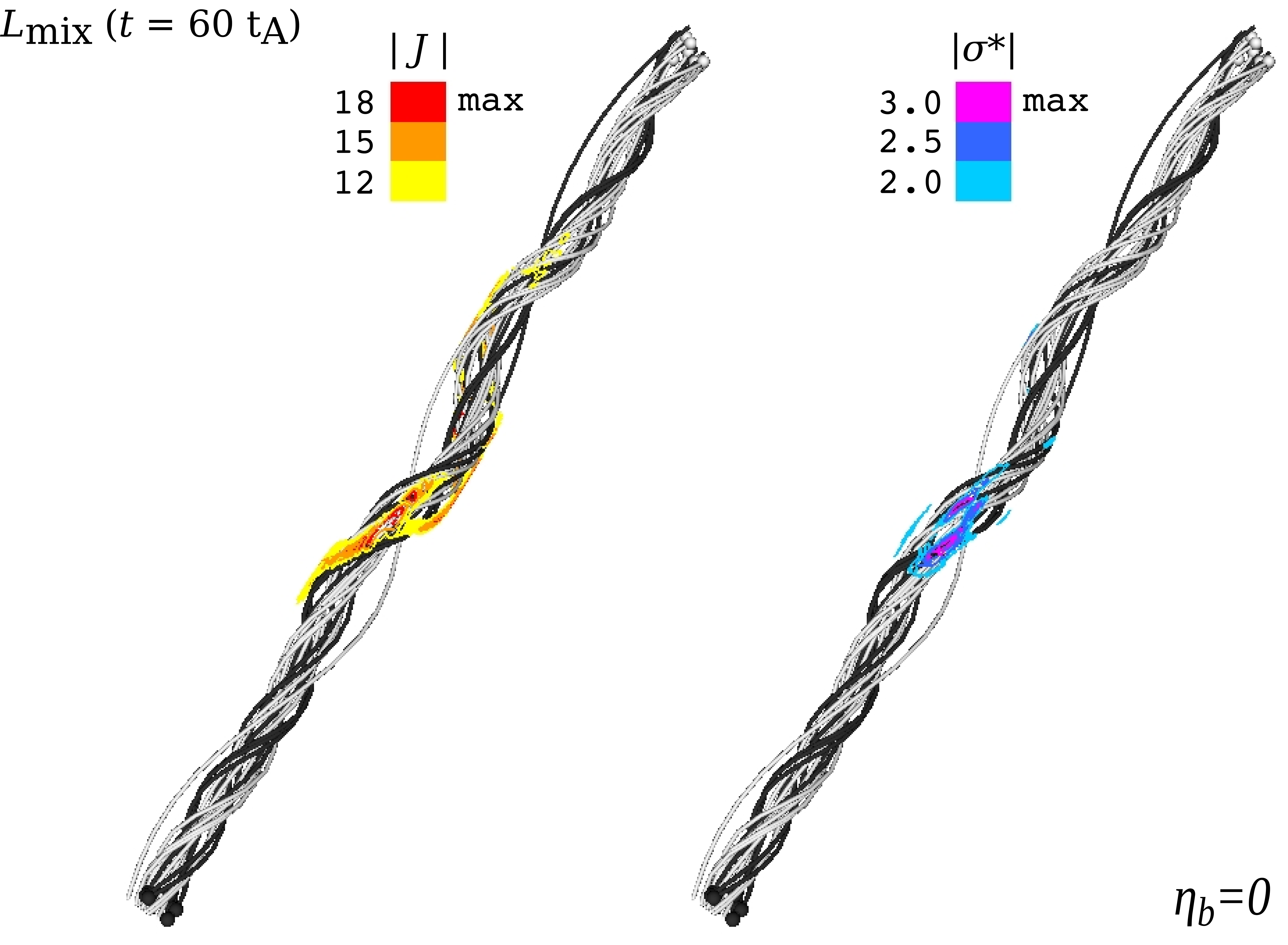}  
  \caption{\small{Magnetic field lines originating from the bottom left footpoint (dark grey) and from the upper right footpoint (light grey) are shown at $t\,{=}\,60\,t_{\mathrm{A}}$ for $L_{\mathrm{uni}}$ (top) and $L_{\mathrm{mix}}$ (bottom). At the onset of instability, two plots are shown: one with isosurfaces of current (left) and the other with isosurfaces of $|\,\sigma^*\,|$, a proxy for viscous heating (right).}}
  \label{b_d_bfield_instability}
\end{figure*}

Now we examine the magnetic field (and critical current distribution) at specific times during the simulations. Figure \ref{b_d_bfield_instability} shows how field lines, originally located within the core, become kinked as the instability takes hold. The dark grey field lines are drawn from the bottom boundary (or footpoint) and the light grey field lines are drawn from similar locations at the upper boundary. Loops $L_{\mathrm{uni}}$ and $L_{\mathrm{mix}}$ follow the same course of events. Initially, the field lines are intertwined; then, during the growth of the instability, the currents become critical (indicated by the yellow, orange and red areas) and anomalous resitivity is applied. The dissipative effects of this anomalous resistivity have a minimal contribution to the internal energy (Section \ref{sec_NumericalResults_EnergyResistivity}); instead magnetic energy is dissipated by the application of viscous heating at shock fronts. Hence, we also show a proxy for viscous heating, $|\,\sigma^*\,|$, which is equivalent to the shock tensor divided by $\,\rho\,L\,(\nu_1\,c_{\mathrm{f}}\,\,+\,\,\nu_2\,L\,|s|)$ --- $|\,\sigma^*\,|$ is represented by the cyan, blue and purple colours. In general, shocks are coincident with current sheets: i.e., the areas of high viscous heating, as indicated by $|\,\sigma^*\,|$, are cospatial with the largest currents. Note, the kink instability is more pronounced for $L_{\mathrm{uni}}$ than for $L_{\mathrm{mix}}$. In terms of azimuthal field, $L_{\mathrm{mix}}$ is the weaker of the two (Figure \ref{znc_b_rx_pf}), however, some importance should be attached to the fact that $L_{\mathrm{mix}}$ is twisted both ways. Opposing twists appear to mitigate the growth of the instability and thereby limit the energy release. If we increase the values of $\alpha_1$ and $\alpha_2$ but keep the opposite signs, such that the total azimuthal field strength is comparable to $L_{\mathrm{uni}}$ (e.g., Loop E), the energy release increases only slightly ($\delta W\,{\approx}\,2.6$).

\begin{figure*}
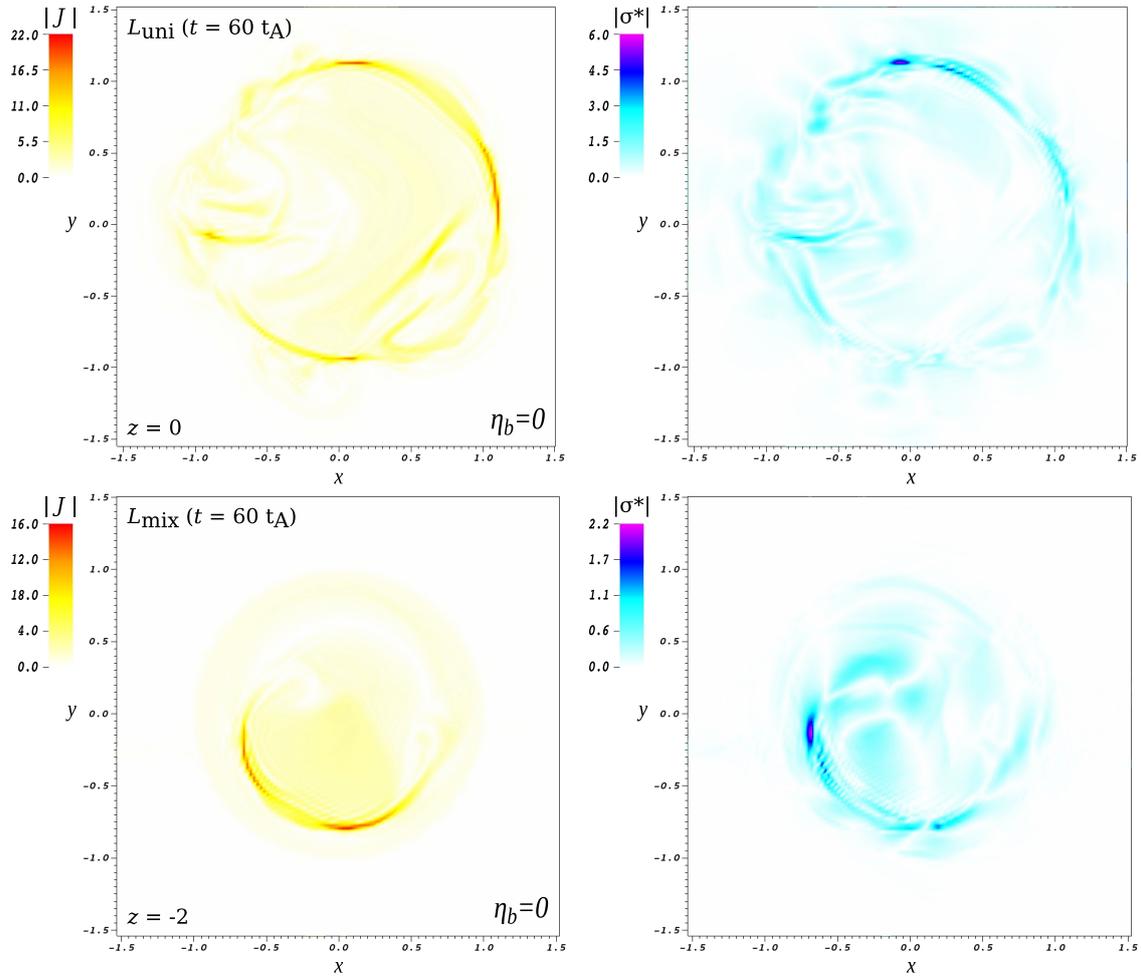

  \center
  \vspace{5pt}
  \includegraphics[scale=0.067]{b_jmag_instability}
  \includegraphics[scale=0.067]{d_jmag_instability}
  \caption{\small{The spatial variation of current magnitude (left) and a viscous heating proxy (right) across the loop cross section at the apex (i.e., where $z\,{=}\,0$) for $L_{\mathrm{uni}}$ (top) and at $z\,{=}\,-2$ for $L_{\mathrm{mix}}$ (bottom).}}
  \label{b_d_midplane_instability}
\end{figure*}

Figure \ref{b_d_midplane_instability} shows cross sections of $L_{\mathrm{uni}}$ at $z\,{=}\,0$ (halfway along the loop) and $L_{\mathrm{mix}}$ at $z\,{=}\,-2$, which is roughly the centre of the only patch of significant viscous heating for $L_{\mathrm{mix}}$ at $t\,{=}\,60\,t_{\mathrm{A}}$, see bottom row of Figure \ref{b_d_bfield_instability}. Again, the colours represent current magnitudes (left column) and viscous heating (right column), and the plot times are the same as those used for Figure \ref{b_d_bfield_instability}; i.e., shortly after the start of the kink instability. At this time, current sheets of narrow width start to form; furthermore, nearly all of these current sheets are associated with shock formation and viscous heating.

\begin{figure*}
  \vspace{5pt}
  \center
  \includegraphics[scale=0.067]{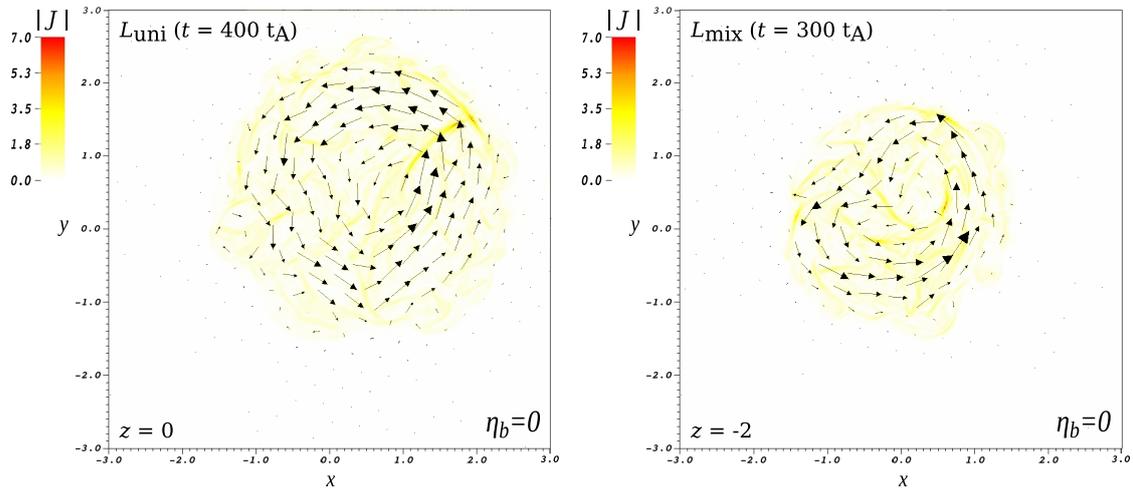}
  \caption{\small{The spatial variation of current magnitude across the loop cross section at the final time of the simulation for $L_{\mathrm{uni}}$ (left) and $L_{\mathrm{mix}}$ (right). All currents are now well below the critical value. The arrows are magnetic field vectors.}}
  \label{b_d_midplane_final}  
\end{figure*}

The unstable phase is over quickly ($\Delta{t}\,{\approx}\,50\,t_{\mathrm{A}}$) and by the end of the simulation the (reconnected) field lines have straightened considerably, indicative of a low constant-$\alpha$ configuration. The areas where shocks have formed must be dispersed throughout the loop volume in order for helicity to be more evenly redistributed and thereby create a linear $\alpha$-profile. The reduction in the azimuthal components of the field lines should cause a radial expansion of the loop. At the initial equilibrium, the inward tension force of the azimuthal field is balanced by the outward magnetic pressure of the axial field; thus, if the tension decreases, the loop must expand before equilibrium can be regained. This behaviour is clearly demonstrated by Figure \ref{b_d_midplane_final}; note the change in scale for the $x$ and $y$ axes. The expansion of the loop is mostly associated with the reconnection of initially twisted field lines inside the loop with the ambient axial field. In Figure \ref{b_d_midplane_final}, the final state calculated by the numerical simulation is overlaid with magnetic field vectors in the $x$-$y$ plane, which are consistent with a cylindrical configuration, bounded by a current-neutralising layer: the arrows follow each other and the arrow sizes initially increase away from the axis, and then diminish before the loop edge. 

Loop $L_{\mathrm{mix}}$ expands less than $L_{\mathrm{uni}}$, which is possibly due to the fact that the latter releases more energy. A single weakly-twisted flux tube best describes the final state for both loops. The following sections will show that the properties of these loops are consistent with relaxation theory.

\subsection{Helicity conservation}
\label{sec_NumericalResults_HelicityConservation}
DeVore (2000) showed how to calculate the magnetic helicity over an entire coronal volume above a photospheric bounding surface. The first step is to work out the magnetic vector potential for a current-free field that has the same distribution of vertical magnetic flux at the lower boundary. DeVore begins by deriving an expression for the scalar potential, using Green's function for Laplace's equation as the integration kernel,
\begin{eqnarray}
  \label{scalar_potential}
  \phi_{\mathrm{c}}(x,y,z',t) & = & \frac{1}{2\pi}\int_{-3}^{+3} dx'\,\,\int_{-3}^{+3} dy'\,\,\frac{B_{z}\big(x',y',-10,t\big)}{r'}\,,
\end{eqnarray}
where $r'\,{=}\,\sqrt{(x-x')^2 + (y-y')^2 + (z')^2}$. The grid domain used by LARE3D has a Cartesian geometry: the coronal loop is initially represented as a straight cylinder within a rectangular box. The $x$ and $y$ axes extend between ${-}3$ and ${+}3$; hence, the integral limits given above. The photospheric boundaries are located at the limits of the $z$ axis ($z\,{=}\,{-}10,\,{+}10$) and $z\,{=}\,0$ is the loop apex; Eq. (\ref{scalar_potential}) uses the first boundary position. The vector potential is constructed using,
\begin{eqnarray}
  \label{vector_potential}
  \vec{A}_{\mathrm{c}}(x,y,z,t) & = & \nabla\times\hat{z}\int_{z}^{+10} dz'\,\,\phi_{\mathrm{c}}(x,y,z',t)\,;
\end{eqnarray}
which becomes,
\begin{eqnarray}
  \label{vector_potential_expanded}
  \nonumber \vec{A}_{\mathrm{c}}(x,y,z,t) & = & \frac{1}{2\pi}\int_{z}^{+10} dz'\int_{-3}^{+3} dx'\int_{-3}^{+3} dy'\\
  &  & \times\,\,\frac{B_{z}\big(x',y',-10,t\big)}{(r')^{3}}\Big[\big(x-x'\big)\,\hat{y}\,-\,\big(y-y'\big)\,\hat{x}\,\Big]
\end{eqnarray}
when the derivatives are moved inside the integral. Now the gauge-invariant vector potential can be specified as,
\begin{eqnarray}
  \label{gauge_invariant_vector_potential}
  \vec{A}\big(x,y,z,t\big) & = & \vec{A}_{\mathrm{c}}\big(x,y,z,t\big)\,-\,\hat{z}\times\int_{-10}^{\,z} dz'\,\,\vec{B}\big(x,y,z',t\big)\,,
\end{eqnarray}  
by subtracting the helicity due to the potential field, and Eq. (\ref{gauge_invariant_vector_potential}) can be re-expressed by expanding the cross product of the second term,
\begin{eqnarray}
  \label{gauge_invariant_vector_potential_expanded}
  \nonumber \vec{A}\big(x,y,z,t\big) & = & \vec{A}_{\mathrm{c}}(x,y,z,t)\,\,+\,\,\int_{-10}^{\,z} dz'\\
  & & \times\,\,\Big[B_{y}\big(x,y,z',t\big)\,\hat{x}\,\,-\,\,B_{x}\big(x,y,z',t\big)\,\hat{y}\Big]\,.
\end{eqnarray}
Finally, the gauge-invariant magnetic helicity is
\begin{eqnarray}
  \label{gauge_invariant_magnetic_helicity}
  \nonumber K & = & \int_V \vec{A}\,{\cdot}\,\vec{B}\,\,dV\\
  & = &  \int_{-3}^{+3} dx\int_{-3}^{+3} dy\int_{-10}^{+10} dz\,\,\vec{A}\big(x,y,z,t\big)\cdot \vec{B}\big(x,y,z,t\big)\,.
\end{eqnarray}
The geometry used by DeVore differs significantly from that used here (Figure \ref{znc_schematic}), which features two separate photospheric boundaries at the limits of the $z$ axis. Fortunately, the relative positions of the two boundaries mean that if the flux is cancelled for one it will be cancelled for the other, and so the lower bound $z$ coordinate can simply be set to $\,{-}10$.

Equations (\ref{scalar_potential})--(\ref{gauge_invariant_magnetic_helicity}) have been implemented, using the five-point Newton-Cotes integration formula. The marginally unstable loops (Figure \ref{znc_it_a2p_larepts}) have zero net current initially and should continue to do so during the simulation; thus, outside the loop the helicity is zero. This means the helicity, calculated using a straightforward cylindrical geometry, can be compared easily to that calculated for a Cartesian geometry, where the loop is enclosed within a rectangular box. The helicity is zero everywhere in the additional volume between the surface of the rectangular box and the outer edge of a cylindrical potential envelope.
\begin{center}
  \begin{table}[h!]
    \caption{Helicity at three times during the simulations ($\eta_{\mathrm{b}}\,{=}\,0$) for all five kink-unstable loops.}
    \label{helicity_calculations}
    \begin{center}    
    \begin{tabular}{ l  c  c  c  c }
      \hline
      \textbf{Loop} & \textbf{Initial} & \textbf{Instability} & \textbf{Final} & $\Delta{K}/\Delta{W}$ \\ \hline      
      \textbf{A} & 12.3 & 12.26 ($t\,{=}\,50\,t_{\mathrm{A}}$) & 12.22 & 0.09\\ \hline
      \textbf{B ($L_{\mathrm{uni}}$)} & 12.29 & 12.27 ($t\,{=}\,60\,t_{\mathrm{A}}$) & 12.28 & 0.03 \\ \hline
      \textbf{C} & 10.47 & 10.46 ($t\,{=}\,100\,t_{\mathrm{A}}$) & 10.5 & 0.19 \\ \hline
      \textbf{D ($L_{\mathrm{mix}}$)} & 6.12 & 6.12 ($t\,{=}\,60\,t_{\mathrm{A}}$) & 6.11 & 0.13 \\ \hline
      \textbf{E} & 1.16 & 1.14 ($t\,{=}\,50\,t_{\mathrm{A}}$) & 1.18 & 1.32 \\      
      \hline    
    \end{tabular}
    \end{center}
  \end{table}
\end{center}
Table \ref{helicity_calculations} gives the helicities for each loop at three different times. The second time is the time of instability; i.e., when the loop is furthest from equilibrium. The helicity appears to be conserved: it varies little over the course of the simulation. For loops A--D, the helicity varies by less than one percent; whereas for Loop E the helicity increases by ${\sim}\,2\%$.

Next, we calculate the ratio of the helicity variation to the change in magnetic energy, 
\begin{eqnarray}
  \label{helicity_variation}
  \frac{\Delta{K}}{\Delta{W}} & = & \Bigg|\frac{\ln(K_{\mathrm{final}}/K_{\mathrm{initial}})}{\ln(W_{\mathrm{final}}/W_{\mathrm{initial}})}\Bigg|\,;
\end{eqnarray}
both $\Delta{K}$ and $\Delta{W}$ are weighted by initial value. For four of the five loops, $\Delta{K}$ is around one order of magnitude lower than $\Delta{W}$ (Table \ref{helicity_calculations}, fifth column); these results are comparable with Browning et al. (2008). The relationship $\Delta{K}\,{\ll}\,\,\Delta{W}$ implies that magnetic energy dissipation is taking place on small spatial scales (i.e., shock fronts). One of the loops (E) has $\Delta{K}\,{>}\,\Delta{W}$; however, it is no coincidence that this loop also has the lowest helicity ($K_{\mathrm{initial}}\,{=}\,1.16$). The coarseness of the grid sampling used to calculate Eq. (\ref{gauge_invariant_magnetic_helicity}) --- the $x$ and $y$ dimensions are sampled at every fourth cell and the $z$ dimension at every other cell --- is too great for such a small helicity. Hence, for this last loop, the grid sampling is increased from $4{\times}4{\times}2$ to $2{\times}2{\times}2$, which gives $\Delta{K}/\Delta{W}\,{\approx}\,0.44$. A finer sampling and/or higher resolution is required to bring this ratio down to below 0.1.

\section{Partial relaxation model}
\label{sec_PartialRelaxationModel}
\subsection{Analytical calculation}
\label{sec_PartialRelaxationModel_AnalyticalCalculation}
In Browning \& Van der Linden (2003), Browning et al. (2008) and Bareford et al. (2010), a loop with net current was assumed to relax such that it expanded to fill the entire potential envelope: the $\alpha$-profile was invariant between $r\,{=}\,0$ and $r\,{=}\,3$ ($R_{\mathpzc{B}}$). The \textit{relaxed} alpha was identified by assuming that $\psi$ (axial flux) and $K/\psi^2$ (the normalised helicity) were conserved over the loop and envelope, in accordance with Taylor relaxation. The only limit to the relaxation is the position of an (unphysical) bounding wall. Hence, the relaxed state always represented a threefold radial expansion of the initial state. Later, Bareford et al. (2011), considered a range of relaxation radii: $R_{\mathpzc{b}}\,{\leq}\,r\,{\leq}\,R_{\mathpzc{B}}$. Some form of partial relaxation is more likely to be relevant to the zero-net-current case; it is known from simulation results, presented here and in Hood et al. (2009), that reconnection is of limited extent, leaving much of the external field undisturbed. This contrasts with previous results for loops carrying net current (Browning et al. 2008), in which the disturbances in the nonlinear phase of the kink generally extended to the boundary of the simulation.
\begin{figure}[h!]
  \center
  \includegraphics[scale=0.23]{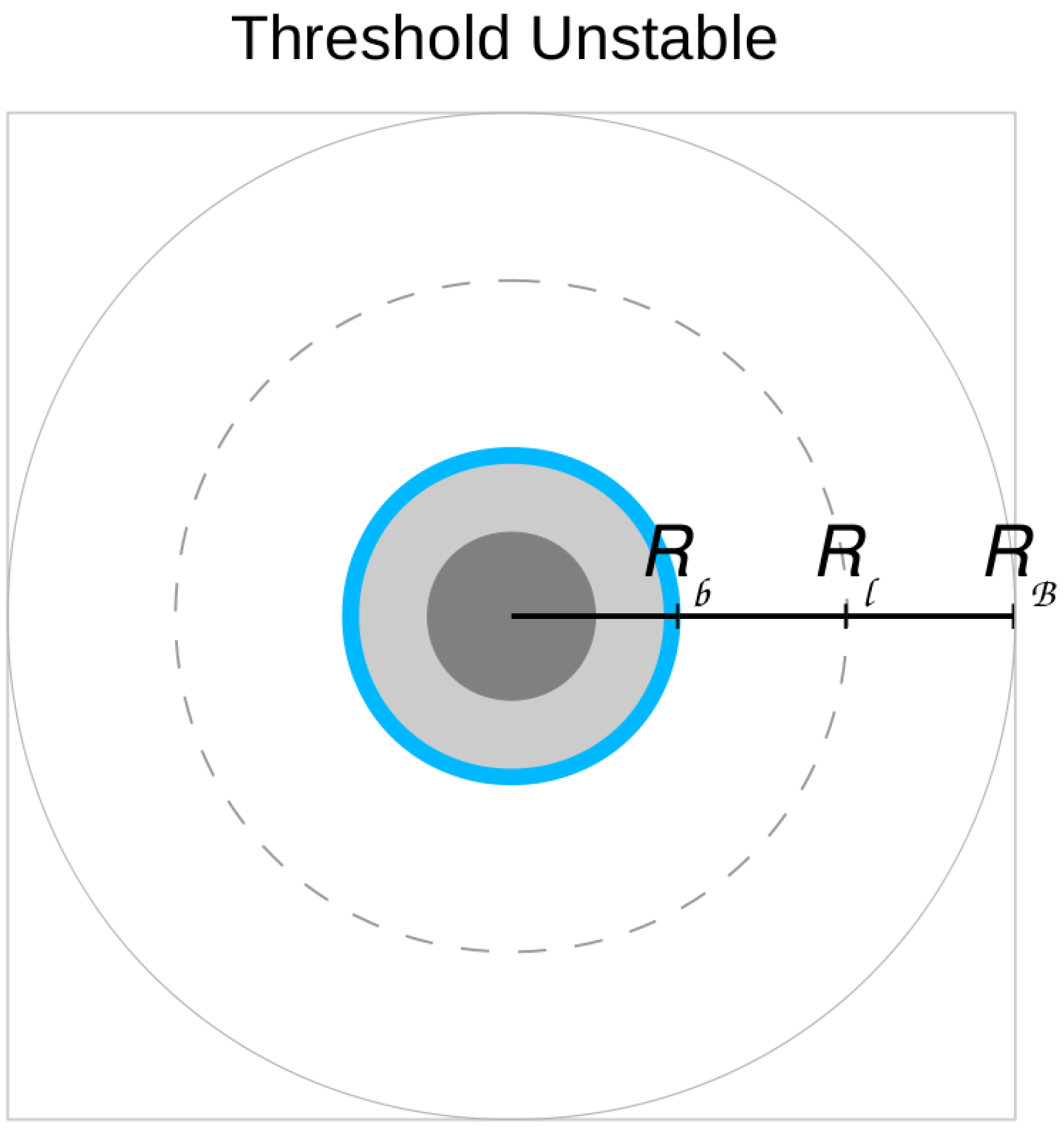}\hspace{4pt}
  \includegraphics[scale=0.23]{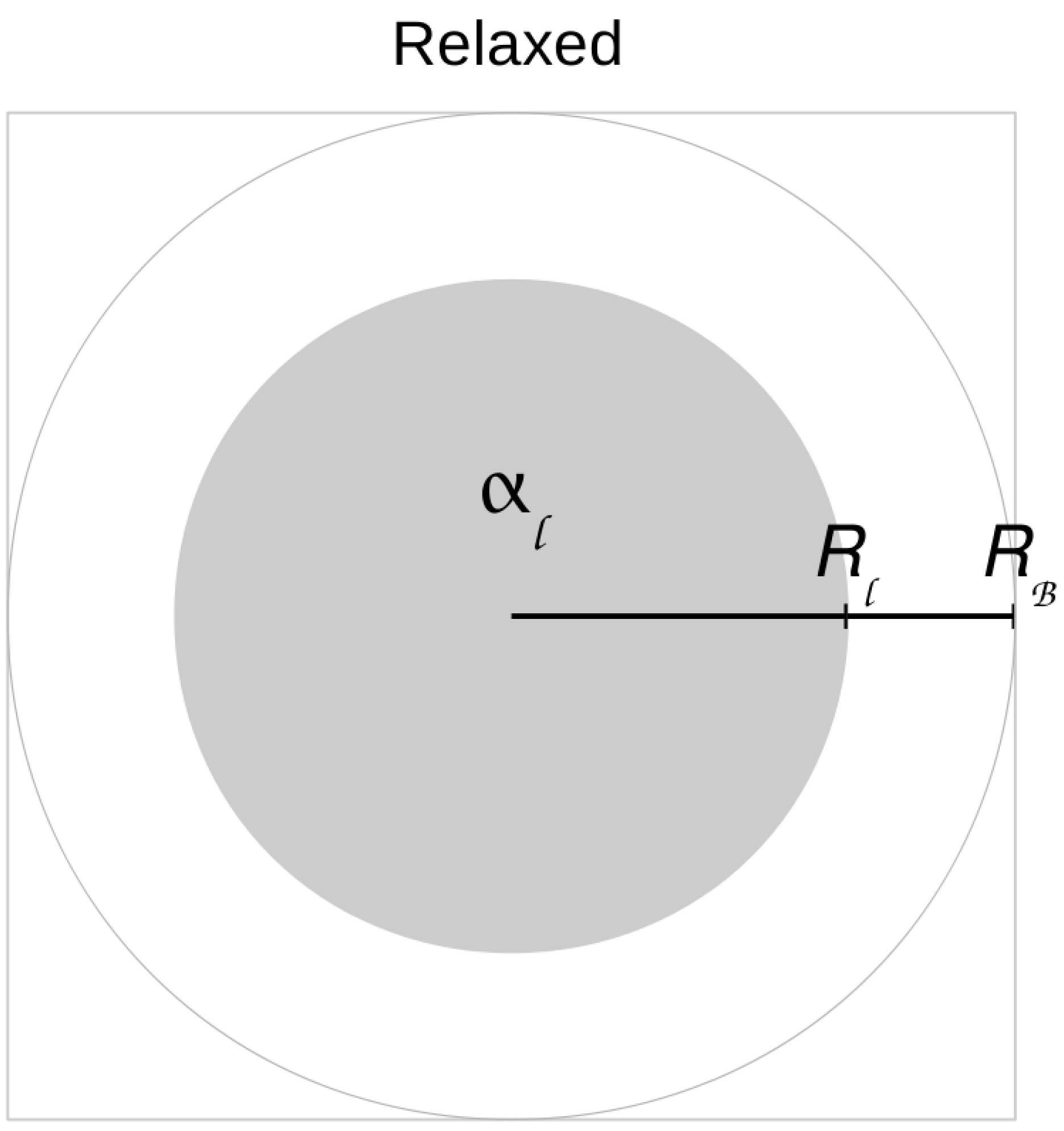}
  \caption{\small{The unstable and relaxed loop. $K/\psi^2$ is conserved over the region enclosed by the dashed circle, which is the outer boundary ($R_{\mathpzc{l}}$) of the relaxed loop.}}
  \label{znc_unstable_to_relaxed}
  \vspace{-5pt}
\end{figure}

Bareford et al. (2011) maintained zero net current by fixing a neutralising loop surface at a specified radius ($r\,{=}\,R_{\mathpzc{l}}$). The neutralising surface is imposed by fixing the field coefficients of the potential envelope, so that they do not change during relaxation. In the relaxed state, the envelope is the region between $R_{\mathpzc{l}}$ and $R_{\mathpzc{B}}$. Naively, it might have been expected that the partially-relaxed state would consist of a constant $\alpha$ field embedded directly in a potential field ($\alpha\,{=}\,0$) with no layer of reversed current. However, such an arrangement does \textit{not} match the simulation results, see Section \ref{sec_PartialRelaxationModel_NumericalRelaxedStates}. Furthermore, it would be unphysical for the partially-relaxed state to include a region of azimuthal field extending to infinity, which is the consequence of allowing net current, since in the initial state, the azimthal field is zero outside the loop. The existence of current sheets (here broadened into a narrow current-neutralising layer) bounding localised relaxed states has also been noted by Gimblett et al. (2006). The axial flux is conserved such that $\psi_{\mathpzc{l}}$ of the threshold state is equal to $\psi_{\mathpzc{l}}$ of the relaxed state. The subscript $\mathpzc{l}$ denotes the relaxation radius, $R_{\mathpzc{l}}$; it is the radial upper bound over which the associated property is calculated --- the lower bound being the axis. For example, $\psi_{\mathpzc{l}}$ is the axial flux from $r\,{=}\,0$ to $r\,{=}\,R_{\mathpzc{l}}$ (similarly, $K_{\mathpzc{B}}$ is the helicity from $r\,{=}\,0$ to $r\,{=}\,R_{\mathpzc{B}}$, the outer edge of the potential envelope). This method of conservation is illustrated by Figure \ref{znc_unstable_to_relaxed}. Left, is the marginally unstable threshold state; the radius that this loop will expand to is indicated by the dashed circle; right, is the expanded relaxed loop. We might expect that, either the relaxed loop flux ($\psi_{\mathpzc{l}}$) matches the initial loop flux ($\psi_{\mathpzc{b}}$), or it matches the initial flux within the radius ($R_{\mathpzc{l}}$) eventually attained. The former is correct if the field freely expands into the surrounding field. The latter choice is made if the loop radially expands by reconnection; i.e., the loop \textit{eats into} the surrounding axial field. This outcome agrees much better with the simulation results (again see Section \ref{sec_PartialRelaxationModel_NumericalRelaxedStates}). If the loop did not reconnect with its surroundings \textit{and} expand to a radius of $1.8\,R_{\mathpzc{b}}$ (as for $L_{\mathrm{uni}}$), then the axial field would drop by a factor of $1.8^2$, which is not observed. (The slight mismatch between the numerical and analytical axial field profiles seen in Figure \ref{znc_cfgb_b} is probably due to some limited free expansion of the loop.)

The relaxation alpha ($\alpha_{\mathpzc{l}}$) is determined by ensuring that the axial flux and helicity integrated over the dashed circle in Figure \ref{znc_unstable_to_relaxed} (left), match the values obtained when these same properties are integrated over $0\,{\leq}\,r\,{\leq}\,R_{\mathpzc{l}}$ of the relaxed loop (Figure \ref{znc_unstable_to_relaxed}, right). Also, helicity is absent from the potential envelope surrounding a zero-net-current loop; hence, $K_{\mathpzc{b}}\,{=}\,K_{\mathpzc{B}}$ (unlike magnetic energy, $W_{\mathpzc{b}}\,{\neq}\,{W}_{\mathpzc{B}}$). Thus, we do not need to choose which value of helicity to use. Once $\alpha_{\mathpzc{l}}$ is known, the energy release can be determined analytically;
\begin{eqnarray}   
  \label{eqn_dw_scn23}\delta W & = & W_{\mathpzc{l}}(\alpha_{\mathrm{i1}},\alpha_{\mathrm{i2}}) - W_{\mathpzc{l}}(\alpha_{\mathpzc{l}})\,,  
\end{eqnarray}
where $W_{\mathpzc{l}}(\alpha_{\mathrm{i1}},\alpha_{\mathrm{i2}})$ is the energy of the threshold state and $W_{\mathpzc{l}}(\alpha_{\mathpzc{l}})$ is the energy of the relaxed state.

\begin{figure}[h!]
  \center
  \includegraphics[width=120pt,height=90pt]{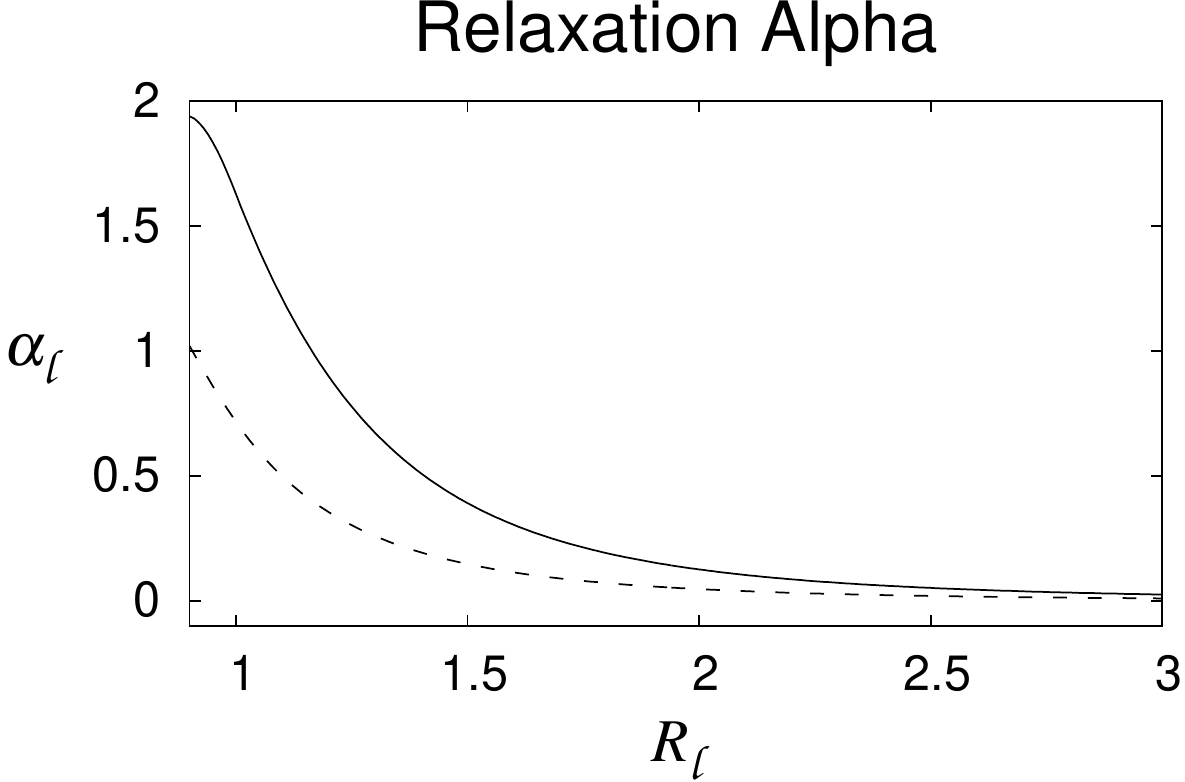}\hspace{10pt}
  \includegraphics[width=115pt,height=90pt]{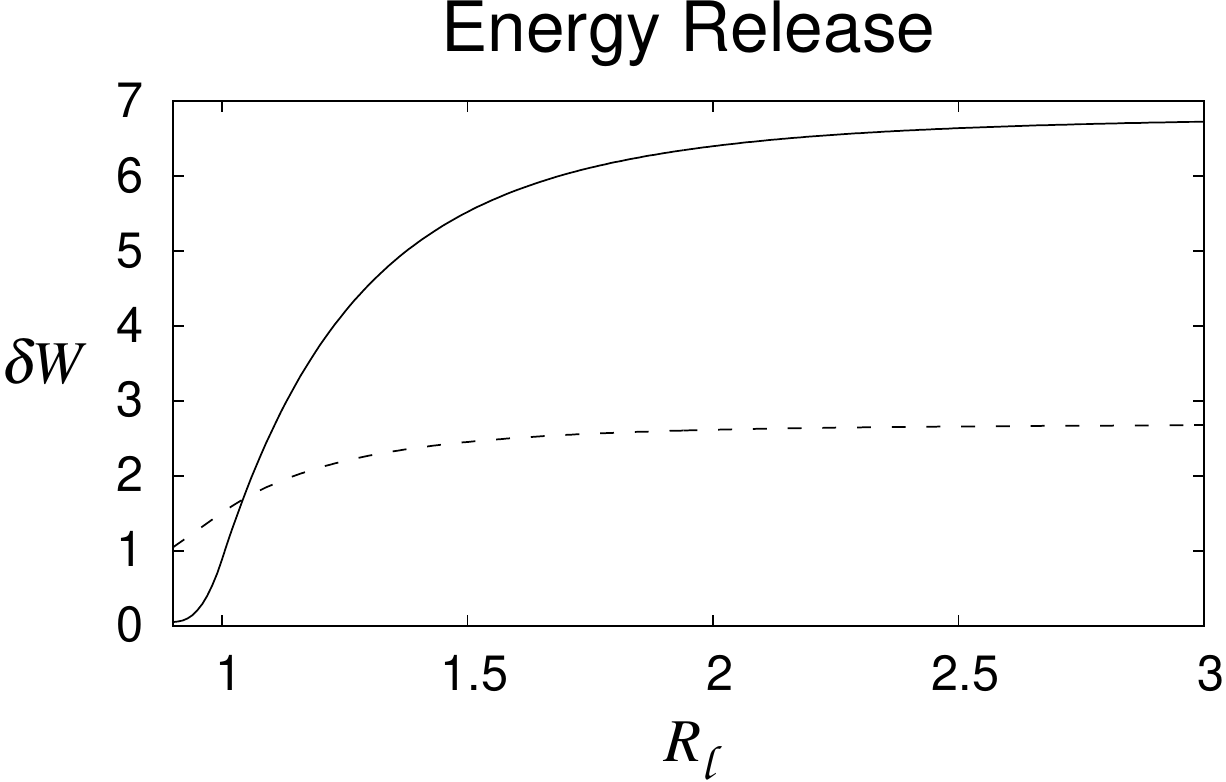}
  \caption{\small{The variation with relaxation radius ($R_{\mathpzc{l}}$) of relaxed alpha ($\alpha_{\mathpzc{l}}$, left) and of dimensionless energy release ($\delta{W}$, right) for $L_{\mathrm{uni}}$ (solid line) and for $L_{\mathrm{mix}}$ (dashed line).}}
  \label{znc_ax_and_wr_with_rx}
\end{figure}
The relaxation radius ($R_{\mathpzc{l}}$) remains as a free parameter, although sensible values can be inferred from the numerical results. Figure \ref{znc_ax_and_wr_with_rx} shows how $\alpha_{\mathpzc{l}}$ and $\delta{W}$ vary with relaxation radius. As expected, these figures show an inverse relationship between $\alpha_{\mathpzc{l}}$ and $\delta{W}$. In general, $\delta{W}$ increases with $R_{\mathpzc{l}}$, however, this relationship is not linear; beyond a moderate expansion of 50\% ($R_{\mathpzc{l}}\,{=}\,1.5$) the energy release is $\sim$99\% of its maximum for $L_{\mathrm{mix}}$ and $\sim$80\% for $L_{\mathrm{uni}}$. This property of diminishing returns is also true for the other three loops indicated in Figure \ref{znc_it_a2p_larepts}. This means that the energy release is insensitive to the choice of relaxation radius, so long as it is assumed that $R_{\mathpzc{l}}\,{\gtrsim}\,1.5$.

\subsubsection{Numerical Relaxed States}
\label{sec_PartialRelaxationModel_NumericalRelaxedStates}
The final numerically-determined state is merely the last snapshot provided by the simulation; it is expected to be close to the analytically-determined relaxed state.
\begin{figure*}
  \center
  \includegraphics[scale=0.46]{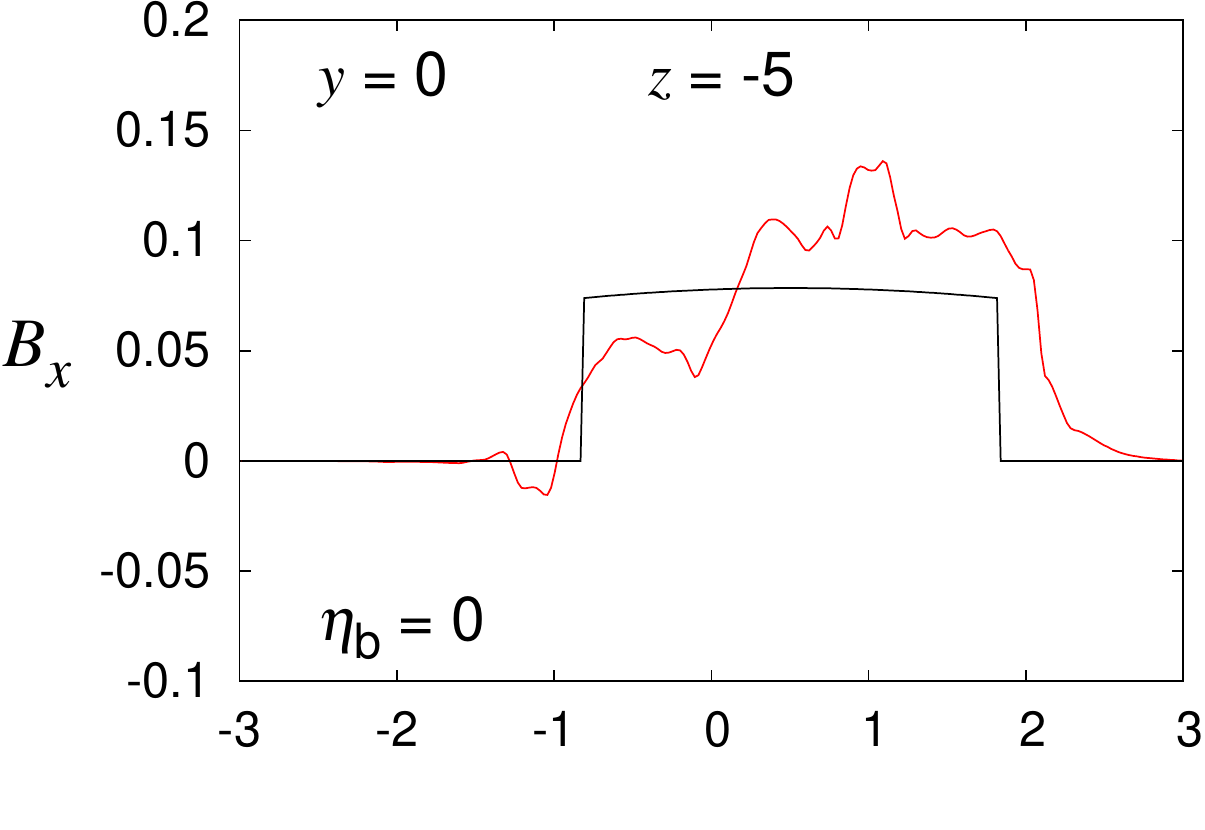}\hspace{5pt}
  \includegraphics[scale=0.46]{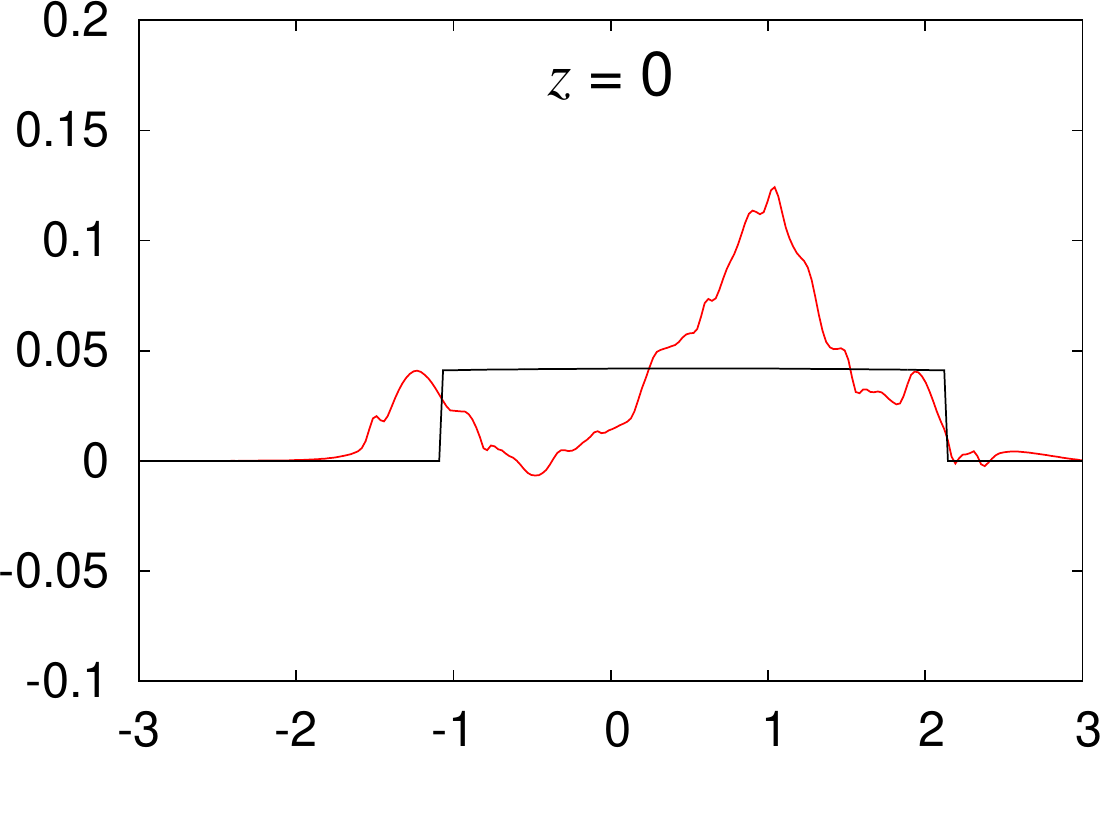}\hspace{5pt}
  \includegraphics[scale=0.46]{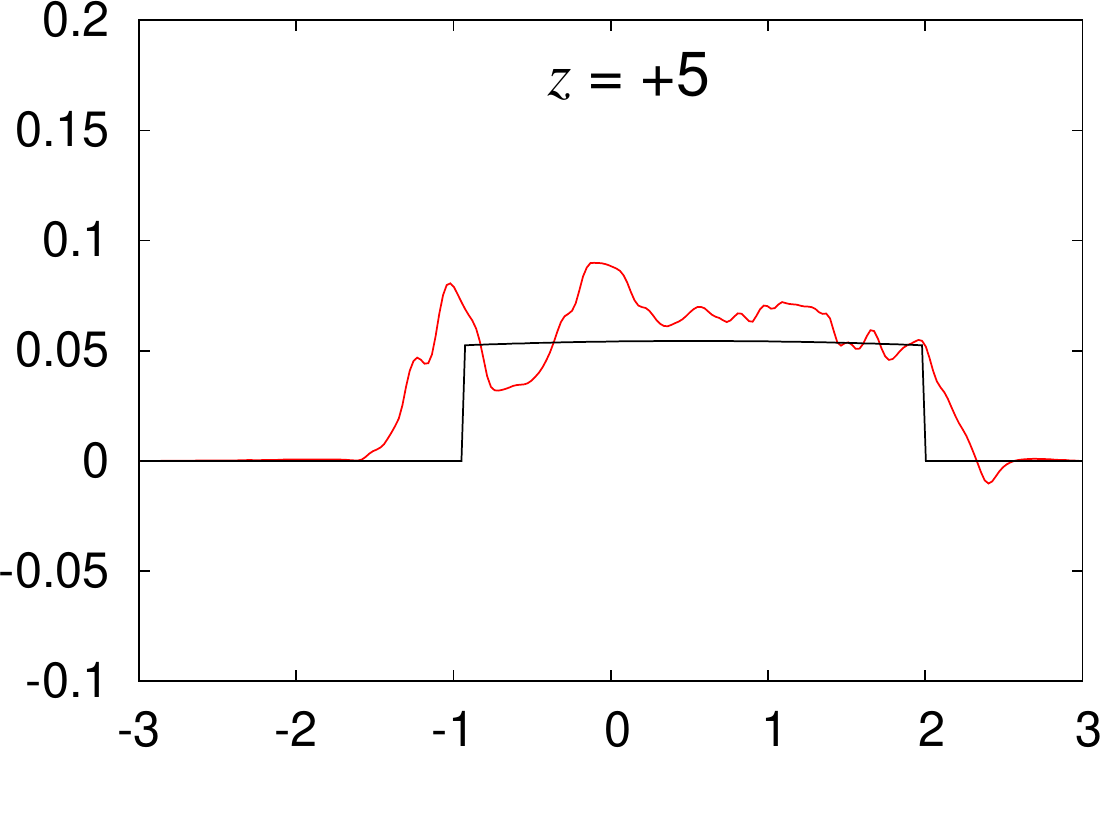}
  \includegraphics[scale=0.46]{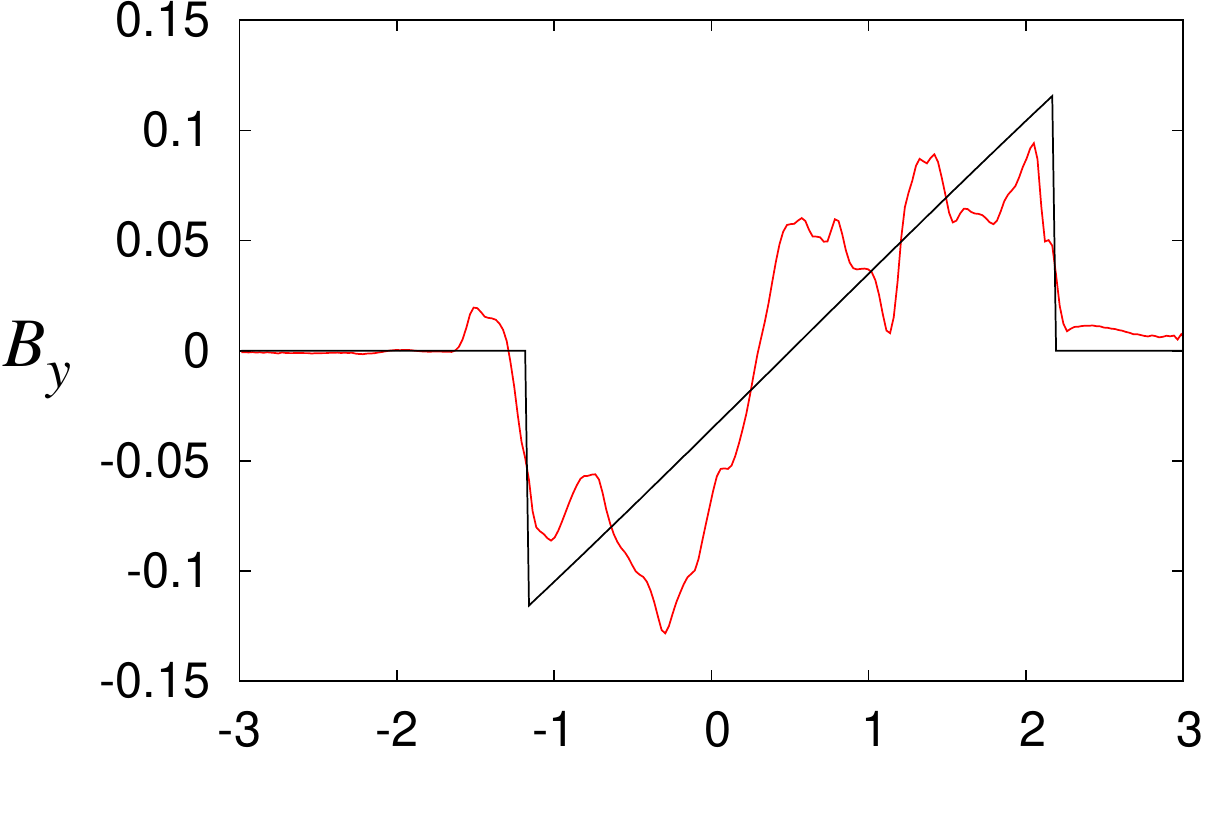}\hspace{5pt}
  \includegraphics[scale=0.46]{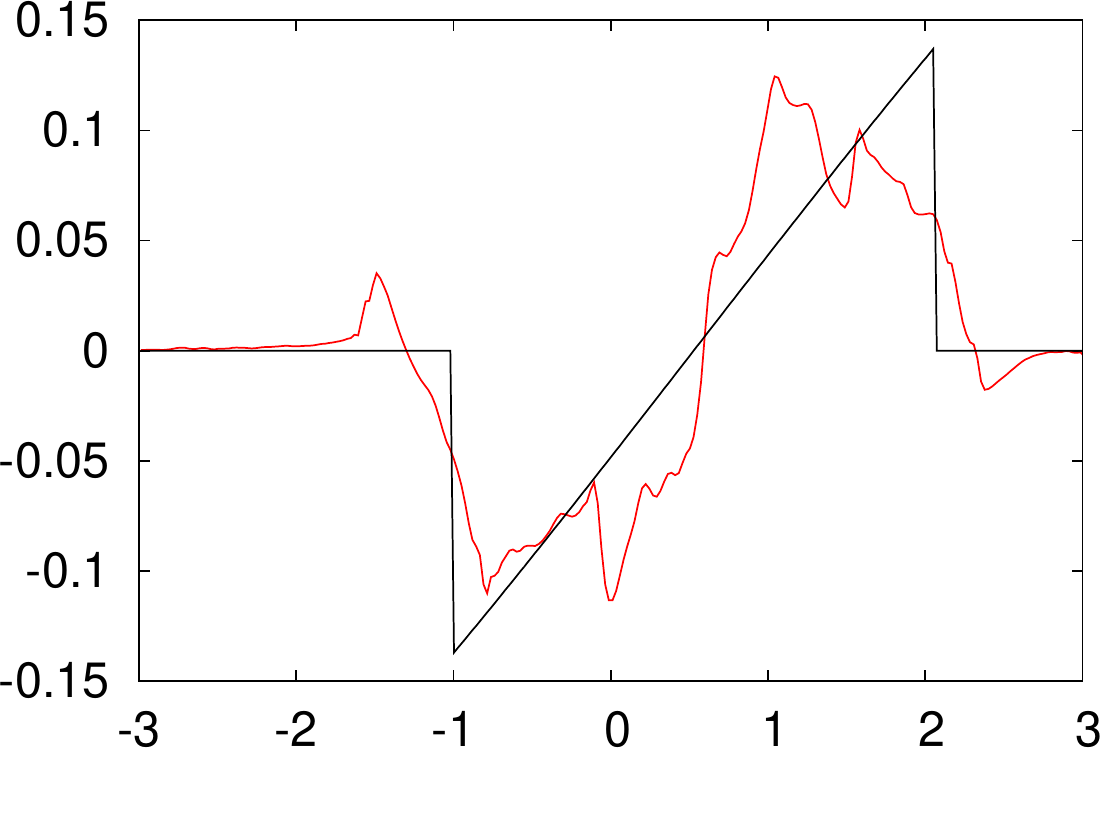}\hspace{5pt}
  \includegraphics[scale=0.46]{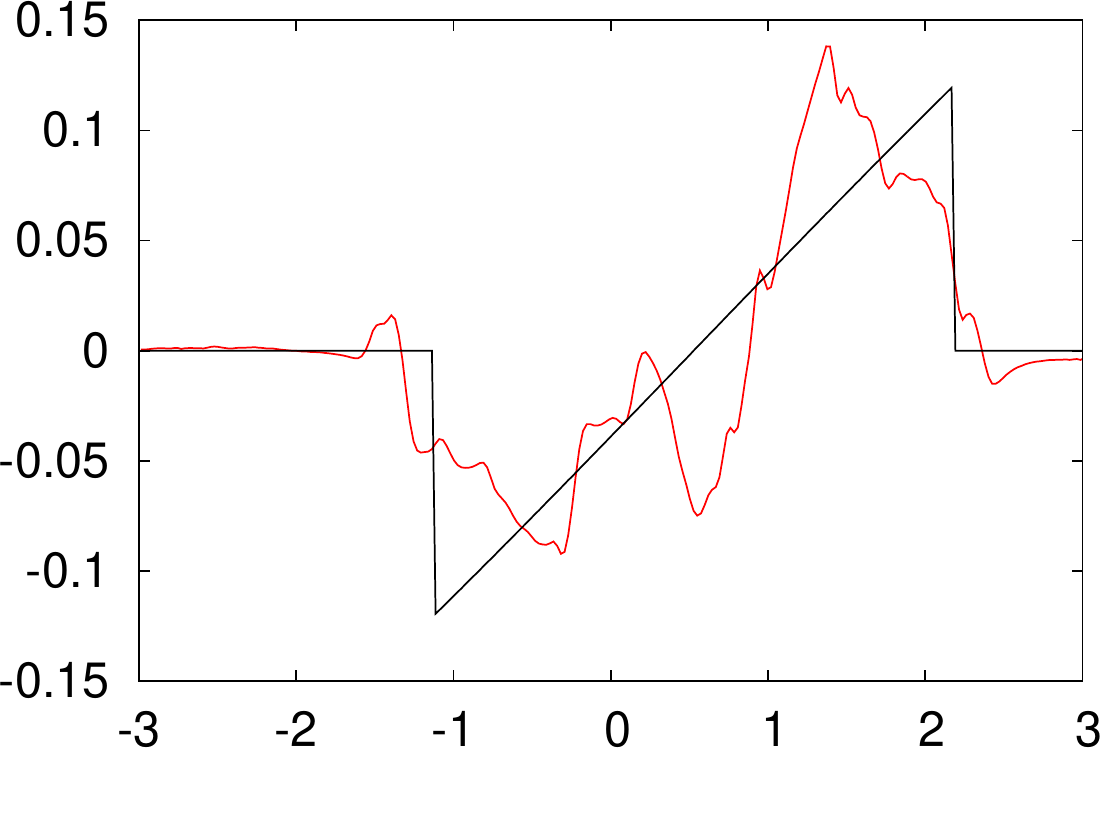}
  \includegraphics[scale=0.465]{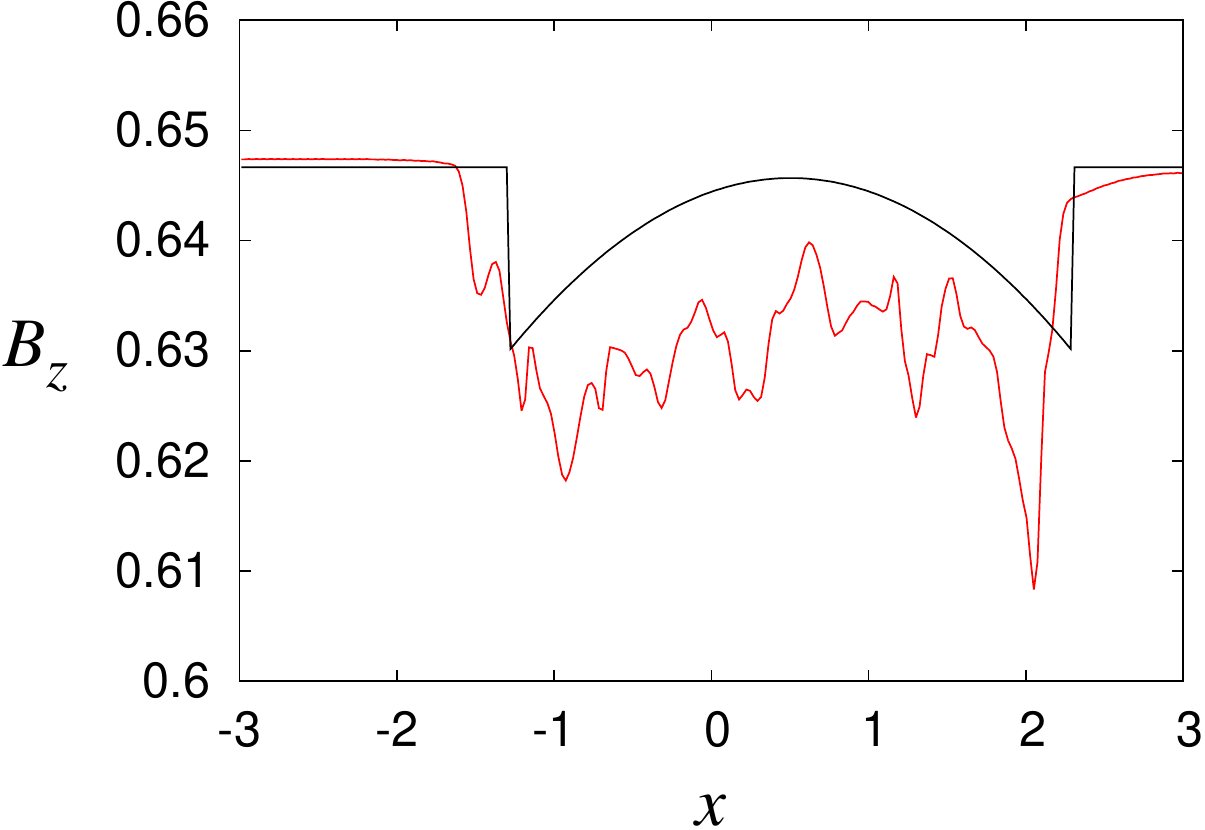}\hspace{5pt}
  \includegraphics[scale=0.465]{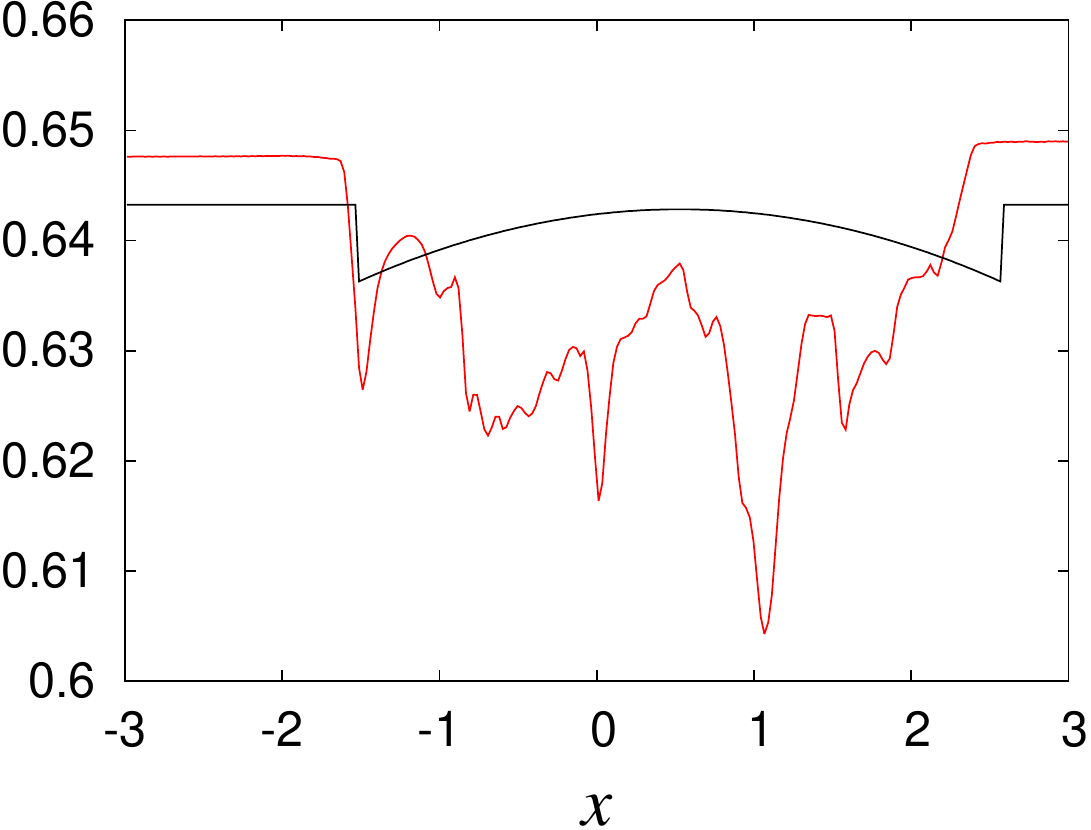}\hspace{5pt}
  \includegraphics[scale=0.465]{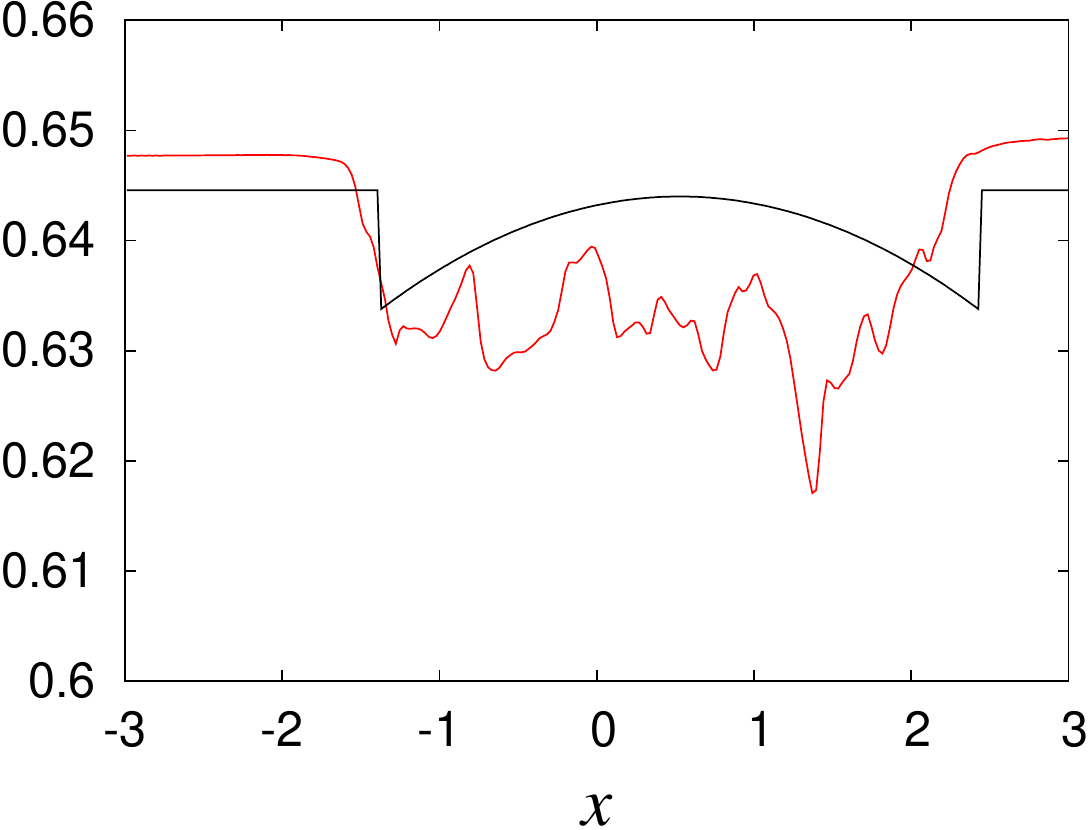}
  \caption{\small{Loop $L_{\mathrm{uni}}$: a comparison between the $B_{x}$ (top row), $B_{y}$ (middle row) and $B_{z}$ (bottom row) magnetic field profiles obtained numerically (red line) and analytically (black line). The latter is calculated from the $\alpha_{\mathpzc{l}}$ and $R_{\mathpzc{l}}$ that best fit the numerical plot, which is taken from the final frame ($t\,{=}\,400\,t_{\mathrm{A}}$) of the high resolution LARE3D simulation ($\eta_{\mathrm{b}}\,{=}\,0$). The comparisons are done at $y\,{=}\,0$ for different $z$ coordinates, $z\,{=}\,-5$ (left column), 0 (middle column) and 5 (right column).}}  
  \label{znc_cfgb_b}
\end{figure*}
Each loop is simulated for at least $300\,t_{\mathrm{A}}$; so, the sooner the instability occurs, the closer the loop will be to a fully relaxed state by the end of the simulation. The Cartesian components of the analytical relaxed field are as follows,
\begin{eqnarray}
  B_{x}(r) & = & -B_{\mathrm{\theta}}(r)(y/r)\,,\\
  B_{y}(r) & = & B_{\mathrm{\theta}}(r)(x/r)\,,\\
  B_{z}(r) & = & B_1 J_0(|\alpha_{\mathpzc{l}}|r)\,,
\end{eqnarray}
where $B_{\theta}(r)\,{=}\,(\alpha_{\mathpzc{l}}/|\alpha_{\mathpzc{l}}|)\,B_1 J_1(|\alpha_{\mathpzc{l}}|r)$ and $r\,{\leq}\,R_{\mathpzc{l}}$. We generate relaxed configurations for every value of $R_{\mathpzc{l}}$ between 0.9 and 3.0 in increments of 0.01. Then, for each field component, it is checked which of the 211 possibilities has the lowest chi-squared value when compared with the numerical values for the same component. These field comparisons ($B_{x}$, $B_{y}$ and $B_{z}$) are performed over the $x$ dimension for a selection of fifteen $y\text{-}z$ coordinate pairs ($y\,{\in}\,\{-1,-0.5,0,+0.5,+1\}$, $z\,{\in}\,\{-5,0,+5\}$) --- the horizontal dashed lines of Figure \ref{znc_cfgb_rx_alpha} (left) indicate the $y$ positions of the field plots. 

As a result of the kink instability, the loop axis will be shifted from the origin in the $x\text{-}y$ plane. The new axis position will be $z$-dependent and can be recalculated by assuming that the relaxed loop is current neutralised (Figure \ref{b_d_midplane_final}): moving inwards from the edge of the envelope, the loop boundary is detected whenever the $\alpha$-value rises above some minimal value. Any axis shift is taken into account before the analytical results are compared with the numerical data. Figure \ref{znc_cfgb_b} presents a subset of the analytical-numerical comparisons for $L_{\mathrm{uni}}$ at $y\,{=}\,0$.

\begin{figure}
  \center
  \includegraphics[scale=0.64]{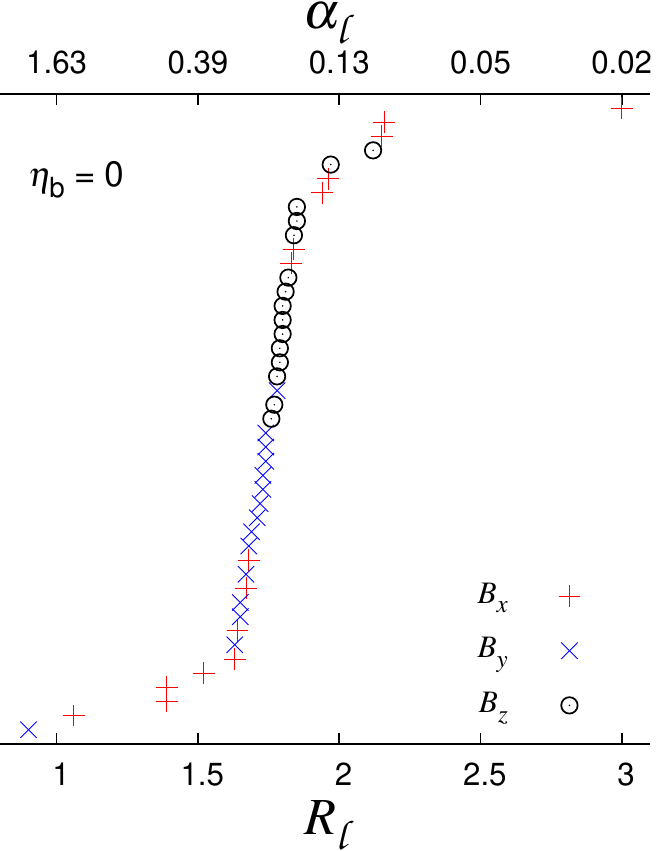} 
  \hspace{5pt}
  \includegraphics[scale=0.64]{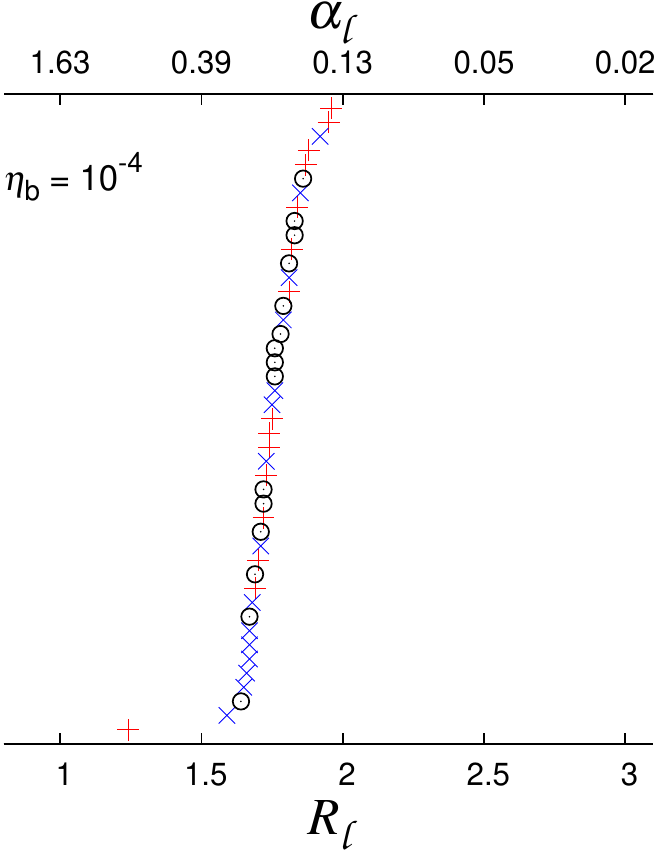}
  \caption{\small{Loop $L_{\mathrm{uni}}$: the best-fit $R_{\mathpzc{l}}\,{\text{-}}\,\alpha_{\mathpzc{l}}$ pairs when $\eta_{\mathrm{b}}\,{=}\,0$ (left) and when $\eta_{\mathrm{b}}\,{=}\,10^{-4}$ (right). Black circles are for those best fits determined from $B_{z}$ profiles, red plus signs are for $B_{x}$ and blue crosses $B_{y}$.}}
  \label{znc_cfgb_rx_avs}
\end{figure}

In general, the analytical plots, such as the ones in Figure \ref{znc_cfgb_b}, show a good agreement with the numerics; however, the $R_{\mathpzc{l}}$ and $\alpha_{\mathpzc{l}}$ that best fit the red lines are independently chosen for each of the forty-five subplots (there are fifteen $y\text{-}z$ positions and three field components). Figure \ref{znc_cfgb_rx_avs} (left) shows the variation in these best-fit parameters: each symbol, by virtue of its position, gives the $R_{\mathpzc{l}}{-}\alpha_{\mathpzc{l}}$ pair that best fits the numerical data extracted at specific $y$ and $z$ coordinates within the simulation domain --- the symbol type denotes which field component is being matched. The use of two $x$-axes (one for $R_{\mathpzc{l}}$, the other $\alpha_{\mathpzc{l}}$) means that all forty-five symbols can be clearly distinguished (the $y$-axis has no meaning beyond a simple sorting of the data points).
The derived averages are $\langle{R_{\mathpzc{l}}}\rangle\,{=}\,1.76\,\pm\,0.29$ and $\langle{\alpha_{\mathpzc{l}}}\rangle\,{=}\,0.28\,\pm\,0.31$; these two properties have a one-to-one mapping. The dimensionless energy release is $5.87\,\pm\,1.11$. Despite the large scatter for $\alpha_{\mathpzc{l}}$, the deviation for $\delta{W}$ is comparatively modest, this is because $d(\delta{W})/d(R_{\mathpzc{l}})$ is small when $R_{\mathpzc{l}}\,{\approx}\,1.8$ (Figure \ref{znc_ax_and_wr_with_rx}, right). Figure \ref{znc_cfgb_rx_avs} suggests that the final numerically-calculated state is tending towards a relaxed loop that can be characterised by a localised invariant $\alpha$-profile. (Note, all points would lie on a single vertical line if there were a perfect match between the analytical and numerical models.) The outlying points on the left have $\alpha_{\mathpzc{l}}$ values that are further from $\langle{\alpha_{\mathpzc{l}}}\rangle$ than those on the right, since the relationship between $\alpha_{\mathpzc{l}}$ and $R_{\mathpzc{l}}$ is not linear (Figure \ref{znc_ax_and_wr_with_rx}, left) - this explains the large deviation for the relaxation alpha. Low levels of magnetic field that fluctuate around the zero line tend to be fitted by low $\alpha_{\mathpzc{l}}$ values; whereas values significantly greater than the mean imply that high currents still exist within the final numerical state. Both of these effects can be ameliorated by re-running the simulation with background resistivity; once again, $\eta_{\mathrm{b}}\,{=}\,10^{-4}$. Figure \ref{znc_cfgb_rx_avs} (right) yields $\langle{R_{\mathpzc{l}}}\rangle\,{=}\,1.74\,\pm\,0.11$ and $\langle\alpha_{\mathpzc{l}}\rangle\,{=}\,0.23\,\pm\,0.09$. The mean dimensionless energy release becomes $\langle\delta{W}\rangle\,{=}\,6.06\,\pm\,0.32$. The resistivity smooths out low-level noise and restricts the current, and thereby reduces the deviation associated with the analytical fit.  

\begin{center}  
  \begin{table}[h!]   
    \caption{Analytical and numerical comparison for the kink-unstable loops A--E.}
    \label{analytical_numerical_comparison}
    \begin{center}    
    \begin{tabular}{ l  c  c  c  c  c  c }
      \cline{2-7}
      \multicolumn{1}{c}{} & \multicolumn{4}{c}{\textbf{Analytical}} & \multicolumn{2}{c}{\textcolor{red}{\textbf{Numerical}}} \\
      \cline{1-7}
      \textbf{Simulation} & \textbf{$R_{\mathpzc{l}}$} & \textbf{$\alpha_{\mathpzc{l}}$} & \textbf{$K$} & \textbf{$|\delta{W}|$} & \textcolor{red}{\textbf{$K$}} & \textcolor{red}{\textbf{$|\delta{W}|$}} \\ \hline      
      \textbf{A} & 1.9 & 0.25 & 12.89 & $8.47\,\pm\,0.42$ & \textcolor{red}{12.19} & \textcolor{red}{8.31} \\ \hline
      \textbf{B ($L_{\mathrm{uni}}$)} & 1.76 & 0.28 & 12.92 & $5.87\,\pm\,1.11$ & \textcolor{red}{12.3} & \textcolor{red}{5.8} \\ \hline
      \textbf{C} & 1.8 & 0.2 & 11.03 & $3.7\,\pm\,0.42$ & \textcolor{red}{10.52} & \textcolor{red}{3.15} \\ \hline
      \textbf{D ($L_{\mathrm{mix}}$)} & 1.59 & 0.19 & 6.45 & $2.38\,\pm\,0.28$ & \textcolor{red}{6.14} & \textcolor{red}{2.15} \\ \hline
      \textbf{E} & 1.18 & 0.19 & 1.22 & $2.63\,\pm\,0.11$ & \textcolor{red}{1.15} & \textcolor{red}{2.58} \\
      \hline    
    \end{tabular}
    \end{center}
    \vspace{-10pt}
  \end{table}
\end{center}

The overall impressive level of agreement demonstrated for $L_{\mathrm{uni}}$, also extends to other positions along the instability threshold, see Table \ref{analytical_numerical_comparison}. This table also confirms that the analytically-calculated helicities of Loops A--E are in approximate agreement with the numerical values, which were derived according to the procedure discussed in Section \ref{sec_NumericalResults_HelicityConservation}. The correspondence between the numerical and analytical energy releases is the most significant finding. There is evidence to suggest that this correlation persists even when different settings are used for the LARE3D parameters controlling resistive MHD (Section \ref{sec_NumericalResults_EnergyResistivity} and Figure \ref{b_energy_heating_jmax_plots}). In addition, these results are consistent with previous work (Browning et al. 2008; Hood et al. 2009). 

\subsection{Final Current Distribution}
\label{sec_PartialRelaxationModel_FinalCurrentDistribution}
The jaggedness of the numerical field profiles shown in Figure \ref{znc_cfgb_b} clearly indicates that there are deviations from a simple locally-constant $\alpha$ profile. 
\begin{figure}[h!]
  \center
  \includegraphics[scale=0.54]{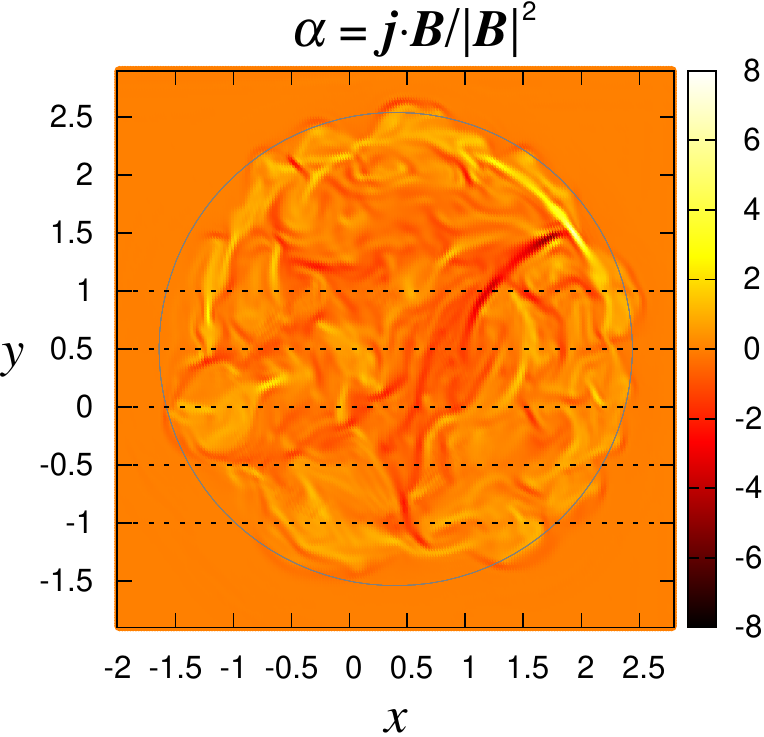}
  \hspace{5pt}
  \includegraphics[scale=0.54]{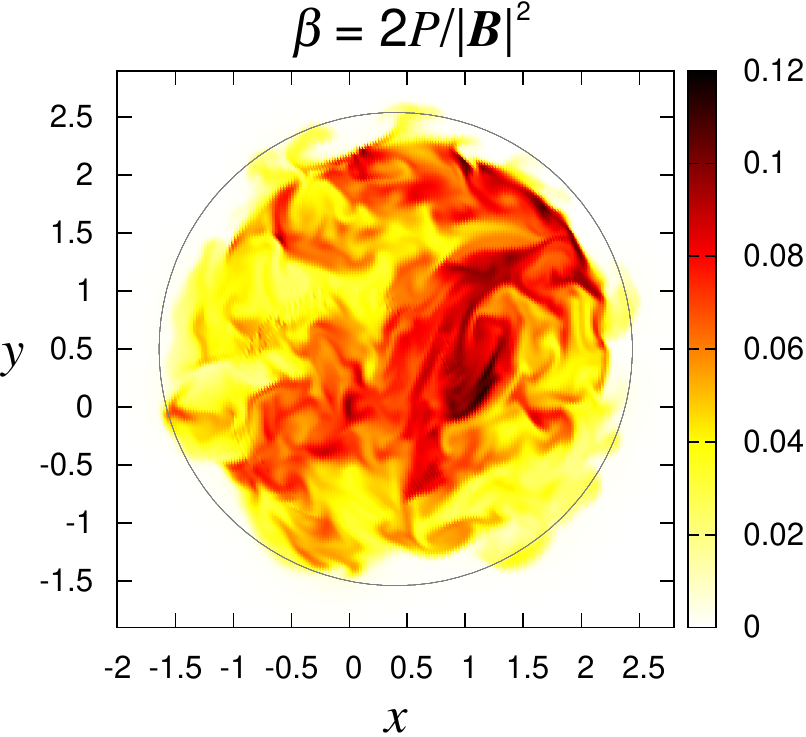}
  \caption{\small{The value of $\alpha$ (left) and \textit{plasma}-$\beta$ (right) over the $x$-$y$ plane for $L_{\mathrm{uni}}$ ($\eta_{\mathrm{b}}\,{=}\,0$). The grey circle approximates the loop cross section at $z\,{=}\,0$ and $t\,{=}\,400\,t_{\mathrm{A}}$.}}
  \label{znc_cfgb_rx_alpha}
\end{figure}
We show the value of $\alpha$ computed over the midplane ($z\,{=}\,0$) of $L_{\mathrm{uni}}$, see Figure \ref{znc_cfgb_rx_alpha}. A grey circle based on the current magnitude plot of Figure \ref{b_d_midplane_final} (left) has been used to approximate the shape of the cross section. There are many, albeit confined, areas that are far from the calculated mean, $\alpha_{\mathpzc{l}}\,{=}\,0.28$. The influence of the initial $\alpha_3$ (the $\alpha$-value for the current neutralisation layer) can be seen in the limits of the colour bar. In addition, \textit{plasma}-$\beta$ (Figure \ref{znc_cfgb_rx_alpha}, right) has, compared to initial values of $10^{-3}$, undergone localised increases of up to two orders of magnitude.

\begin{figure}[h!]
  \center
  \includegraphics[scale=0.53]{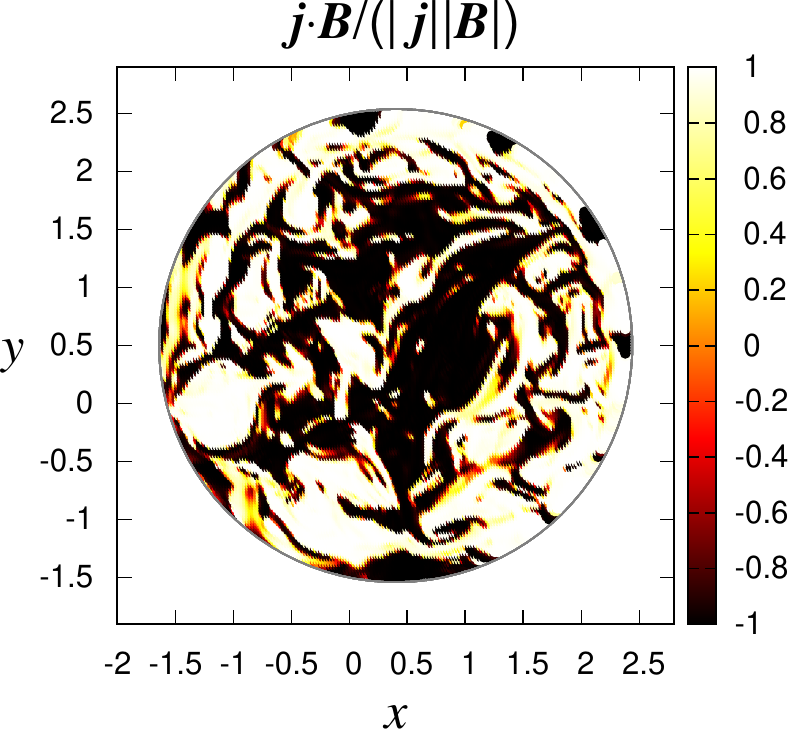}
  \hspace{5pt}
  \includegraphics[scale=0.53]{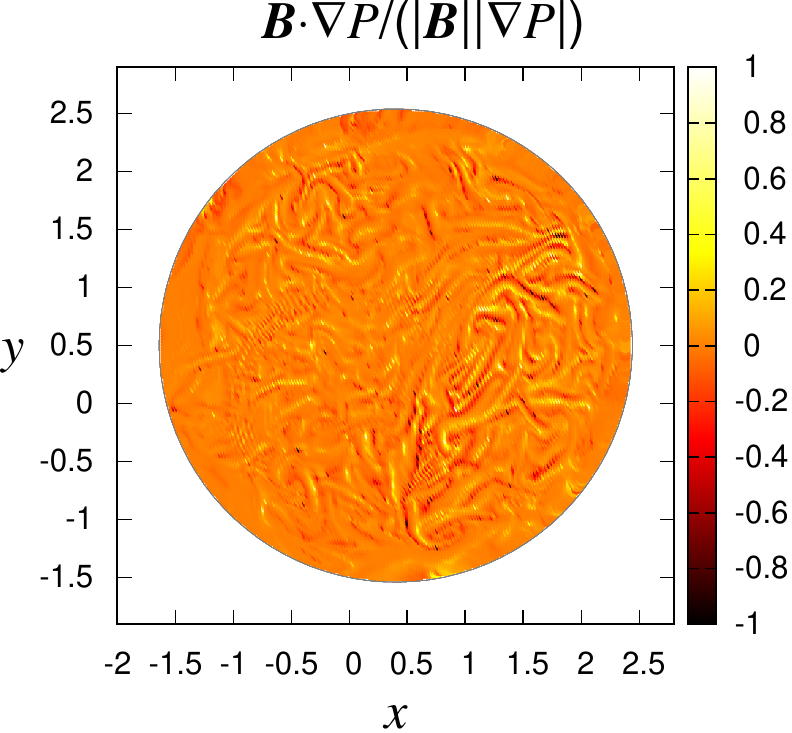}
  \caption{\small{\textbf{Left}, the cosine of the angle between the current and the magnetic field over the $x$-$y$ plane. \textbf{Right}, the cosine of the angle between the magnetic field and the gradient of the pressure over the same area. Positions outside the grey circle, representing the loop cross section, have \textit{not} been assigned a colour. The background resistivity is zero.}}
  \label{znc_cfgb_rx_job_bogp}
  \vspace{-5pt}
\end{figure}
The cosine of the angle between the current and the magnetic field is plotted in Figure \ref{znc_cfgb_rx_job_bogp}; note, positions outside the loop are left unplotted. The current is either parallel (white) or anti-parallel (black) to the field for a high percentage of the cross sectional area (positions that are far from parallel, $20^{\circ}\,{<}\,\vec{j}\angle\vec{B}\,{<}\,160^{\circ}$, account for approximately 30\% of the loop cross section). Almost three quarters of the energy released from the field becomes internal energy (i.e., thermal pressure), which may be associated with departures from a force-free state (although Taylor relaxation would predict a uniform pressure distribution). However, it seems that the loop is heading towards an approximate balance of forces ($\vec{j}\,{\times}\,\vec{B}\,{=}\,\nabla P$): Figure \ref{znc_cfgb_rx_job_bogp} (right) shows that the angle between the field and the pressure gradient is on average close to zero. Interestingly, the parallel and anti-parallel areas are often found next to each other and therefore might be expected to cancel at later simulation times. The plots of Figures \ref{znc_cfgb_rx_alpha} and \ref{znc_cfgb_rx_job_bogp} are qualitatively similar to those taken at other $z$ coordinates that are not too close to the footpoints (e.g., $-5\,{<}\,z\,{<}\,5$).

\section{Summary and conclusions}
\label{sec_SummaryConclusions}
A nonlinear 3D MHD code has been used to simulate the evolution of a set of zero-net-current cylindrical loops. These loop configurations have been identified by a linear analysis as being marginally kink unstable (Bareford et al. 2011). The simulations show that the instability quickly enters a nonlinear phase and magnetic energy declines sharply before leveling off. Furthermore, the amount of energy released matches the amount predicted by Taylor relaxation (Table \ref{analytical_numerical_comparison}), taking account of the fact that the relaxation is localised. Evidence for helicity conservation was presented, and the change in helicity was shown to be much smaller than the drop in magnetic energy. The implication of this result is that energy diffusion is occuring on small scales compared to the global length scale; i.e., within shocks associated with magnetic reconnection. The low values of kinetic energy during the unstable phase (compared to the internal energy) imply that these shocks occur near to reconnection sites. (We find that dissipation within current sheets only becomes significant when $\eta_{b}\,{=}\,10^{-4}$.) Further research could reveal exactly how magnetic reconnection spreads through the loop volume in response to a kink instability, which would also reveal where the loop is heated and when. It is this widespread dispersal of reconnection and shock heating that ensures helicity conservation. At present, we cannot confirm the nature of the shocks: slow-mode shocks are expected since only shocks of this type can reduce magnetic field strength.

Relaxation theory also predicts that the final relaxed state should have a constant $\alpha$-profile. Although the final numerical $\alpha$-profile still retains much fine structure, the final magnetic fields are well-modelled by a (localised) constant-$\alpha$ profile with some fluctuations superposed --- this suggests the fine-scale structure is self-cancelling (i.e., it integrates out). Furthermore, the energy of the final numerical state is very well matched by the energy of the same constant-$\alpha$ state (the field energy is insensitive to the spikeness of the numerical data). The property of zero net current is retained after the instability. Typically, the loop expands radially, the field reconnecting with that present in the potential envelope. These results justify the choices made by Bareford et al. (2011) regarding the details of the relaxation process. Nevertheless, we were concerned that the thinness of the current neutralisation layer (Figure \ref{znc_schematic}) might have influenced the results. Hence, we also ran a resistive MHD simulation (with $\eta_{\mathrm{b}}\,{=}\,0$) for a zero-net-current equilibrium that possessed a smooth $\alpha$-profile, see Case 3 of Hood et al. (2009). The constant parameter, $\lambda\,{=}\,1.62$, was set such that the equilibrium was on the threshold of instability. We found that the simulation results were similar to those produced by $L_{\mathrm{mix}}$. Numerically, the energy release was 1.2, which was again consistent with the analytically-determined value.

It appears that the assumption of a Taylor-relaxed state, subsequent to a kink instability, has been verified by the work presented here. The relaxation does not extend over the full numerical volume, but over a region of smaller extent (out to a radius $R_{\mathpzc{l}}$, which is less than the full radius, $R_{\mathpzc{B}}$). In this sense, the relaxation is partial. A relaxed state can only be identified if the relaxation radius is known; at present, it is unclear how $R_{\mathpzc{l}}$ can be precisely determined from the field configuration at instability onset. However, the analytical work has revealed that for marginally-unstable loops, the energy release varies little with relaxation radius once $R_{\mathpzc{l}}\,{\ge}\,1.5$ (Figure \ref{znc_ax_and_wr_with_rx}); hence, a calculation of the energy release does not necessarily require a precise prediction for $R_{\mathpzc{l}}$.

Energy release could be limited if the unstable loop attains an equilibrium that is less than fully relaxed (i.e., the $\alpha$-profile remains nonlinear) and still conserves helicity. There is perhaps, for some field configurations, another constraint that decides the relaxed state, such as the topological degree of the field line mapping between the ends of the loop, as investigated by Yeates et al. (2010). They examined two braided magnetic field configurations (one based on the simple pigtail braid and the other more complex). Both configurations underwent turbulent relaxation, leading to a final state that conserved topological degree and was \textit{less} relaxed than that predicted by Taylor theory --- the final state for the pigtail braid featured two flux tubes of opposite twist. Nevertheless, it is possible for the Taylor-relaxed state and the state that preserves topological degree to coincide. In our case the invariants given by Yeates et al. do not provide any extra constraint, making our results consistent with their predictions. This would explain the level of agreement between the LARE3D simulations and Taylor relaxation.

An issue for further research concerns the interaction of convective driving with the relaxation process. The LARE3D code could be used to help resolve this issue. It should be possible to choose a loop configuration (i.e., a set of $\alpha$-parameters) that is just inside the threshold for linear kink instability and then, trigger the instability by applying a pre-determined velocity profile ($v_{\theta}$) at one of the footpoints. Loop curvature has also not been considered. The linear stability codes require an analytical form for the magnetic fields. If a loop is to retain its curvature, it can only be simulated numerically, which means choices have to be made concerning loop parameters (e.g., length, radius and $\alpha$-profile). Usefully, those straightened loop configurations that are kink unstable and are likely to be reached by convective driving have been identified. These configurations could be adapted to include curvature and re-simulated within LARE3D. This would reveal what effect, if any, curvature has on the energy release precipitated by kink instability. Of course, this procedure could also be applied to other improvements; e.g., gravity (with $\rho(z)$), conduction and radiation. However, a feature that improves the realism of the loop model may not be important as regards kink instability and Taylor relaxation. 
\\\\
\small{\textit{Acknowledgements}. We thank the referee for helpful comments and M. R. B. acknowledges financial support from STFC.}


\begin{thebibliography}{99}
\bibitem{Arber1999}Arber, T. D., Longbottom, A. W., \& Van der Linden, R. A. M. 1999, ApJ, 517, 990
\bibitem{Arber2001}Arber, T. D., Longbottom, A. W., Gerrard, C. L., \& Milne, A. M. 2001, J. Comput. Phys., 171, 151
\bibitem{Aschwanden2001}Aschwanden, M. J., \& Acton, L. W. 2001, ApJ, 550, 475
\bibitem{Baty1996}Baty, H., \& Heyvaerts, J. 1996, A\&A, 308, 935
\bibitem{Baty2000}Baty, H. 2000, A\&A, 360, 345
\bibitem{Bareford2010}Bareford, M. R., Browning, P. K., \& Van der Linden, R. A. M. 2010, A\&A, 521, A70
\bibitem{Bareford2011}Bareford, M. R., Browning, P. K., \& Van der Linden, R. A. M. 2011, Sol. Phys., 273, 93
\bibitem{Berger1984}Berger, M. A., \& Field, G. B. 1984, J. Fluid. Mech., 147, 133
\bibitem{Berger1999}Berger, M. A. 1999, Plasma Phys. Control. Fusion, 41, B167
\bibitem{Botha2011}Botha, G. J. J., Arber, T. D., \& Hood, A. W. 2011, A\&A, 525, A96+
\bibitem{Browning1986}Browning, P. K., Sakurai T., \& Priest, E. R. 1986, A\&A, 158, 217
\bibitem{Browning1988}Browning, P. K. 1988, J. Plasma Phys., 40, 263
\bibitem{Browning2003}Browning, P. K., \& Van der Linden, R. A. M. 2003, A\&A, 400, 355
\bibitem{Browning2008}Browning, P. K., Gerrard, C., Hood, A. W., Kevis, R., \& Van der Linden, R. A. M. 2008, A\&A, 485, 837
\bibitem{DeVore2000}DeVore, C. R. 2000, ApJ, 539, 949
\bibitem{Dixon2010}Dixon, A. M., Berger, M. A., Browning, P. K., \& Priest, E. R. 1989, A\&A, 225, 156
\bibitem{Evans1988}Evans, C. R., \& Hawley, J. F. 1988, ApJ, 332, 659
\bibitem{Finn1985}Finn, J. M., \& Antonsen, T. M. 1985, Commun. Plasma Phys. Controlled Fusion, 9, 111
\bibitem{Fletcher2009}Fletcher, L. 2009, A\&A, 493, 241
\bibitem{Gerrard2002}Gerrard, C. L., Arber, T. D., \& Hood, A. W. 2002, A\&A, 387, 687
\bibitem{Gerrard2003}Gerrard, C. L., \& Hood, A. W. 2003, Sol. Phys., 214, 151
\bibitem{Gimlett2006}Gimlett, C. G., Hastie, R. J., \& Helander, P. 2006, Plasma Phys. Control. Fusion, 48, 1531
\bibitem{Haynes2007}Haynes, M., \& Arber, T. D. 2007, A\&A, 467, 327
\bibitem{Heidbrink2000}Heidbrink, W. W., \& Dang, T. H. 2000, Plasma Phys. Control. Fusion, 42, L31
\bibitem{Heyvaerts1984}Heyvaerts, J., \& Priest, E. R. 1984, A\&A, 137, 63
\bibitem{Hood1992}Hood, A. W. 1992, Plasma Physics and Controlled Fusion, 34, 411
\bibitem{Hood2009}Hood, A. W., Browning, P. K., \& Van der Linden, R. A. M. 2009, A\&A, 506, 913
\bibitem{Hudson1991}Hudson, H. S. 1991, Sol. Phys. 133, 357
\bibitem{Melrose1994}Melrose, D. B., Nicholls, J., \& Broderick, N. G. 1994, J. Plasma Phys., 51, 163
\bibitem{Priest2005}Priest, E. W., Longcope, D. W., \& Heyvaerts, J. 2005, ApJ, 624, 1057
\bibitem{Qiu2009}Qiu, J. 2009, ApJ, 692, 1110
\bibitem{Srivastava2010}Srivastava, A. K., Zaqarashvili, T. V., Kumar, P., \& Khodachenko, M. L. 2010, ApJ, 715, 292
\bibitem{Taylor1974}Taylor, J. B. 1974, Phys. Rev. Lett., 33, 1139
\bibitem{Taylor1986}Taylor, J. B. 1986, Rev. Mod. Phys., 58, 741
\bibitem{VanderLinden1991}Van der Linden, R. A. M. 1991, Ph.D. Thesis (Katholieke Universiteit Leuven)
\bibitem{VanderLinden1999}Van der Linden, R. A. M., \& Hood, A. W. 1999, A\&A, 346, 303
\bibitem{VanLeer1997}Van Leer, B. 1997, J. Comput. Phys., 135, 229
\bibitem{Vekstein1993}Vekstein, G. E., Priest, E. R., \& Steele, C. D. C. 1993, ApJ, 417, 781
\bibitem{Velli1997}Velli, M., Lionello, R., \& Einaudi, G. 1997, Sol. Phys., 172, 257
\bibitem{Wilkins1980}Wilkins, M. L. 1980, J. Comput. Phys., 36, 281
\bibitem{Withbroe1977}Withbroe, G. L., \& Noyes, R. W. 1977, Ann. Rev. Astr. Ap., 15, 363
\bibitem{Yeates2010}Yeates, A. R., Hornig, G., \& Wilmot-Smith, A. L. 2010, Phys. Rev. Lett. 105, 8
\bibitem{Zhang2003}Zhang, M., \& Low, B. C. 2003, ApJ, 584, 479
\end{thebibliography}
\end{document}